\documentclass[a4paper, amsfonts, amssymb, amsmath, reprint, showkeys, nofootinbib, twoside,notitlepage,onecolumn,preprintnumbers]{revtex4-2}

\def\prd{{Phys.~Rev.~D}}        
\def\prl{{Phys.~Rev.~Lett.}}    
\usepackage{amsmath,amstext}
\usepackage[T1]{fontenc}
\usepackage{amssymb}
\input{epsf}
\usepackage{graphicx}
\usepackage{ae,aecompl}

\usepackage{hyperref}
\usepackage{amsmath}
\usepackage{amssymb}
\usepackage{mathtools}
\usepackage{bm}
\usepackage{cleveref}
\usepackage{tensor}
\usepackage{braket}
\usepackage{enumitem}
\usepackage{mhchem}
\usepackage{cancel}
\usepackage{amsthm}
\usepackage{upgreek}
\usepackage{mathrsfs}

\usepackage{color}

\newcommand{\spc}{\quad \quad \quad}

\def\be{\begin{equation}}
\def\ee{\end{equation}}
\def\beq{\begin{eqnarray}}
\def\eeq{\end{eqnarray}}

\theoremstyle{definition}

\theoremstyle{theorem}

\theoremstyle{corollary}

\begin{document}
\title{First-order relativistic hydrodynamics with an information current}
\author{L.~Gavassino$^1$, N.~Abboud$^2$, E.~Speranza$^3$   \& J.~Noronha$^2$}
\affiliation{
$^1$Department of Mathematics, Vanderbilt University, Nashville, TN, USA
\\
$^2$Illinois Center for Advanced Studies of the Universe \& Department of Physics,
University of Illinois Urbana-Champaign, Urbana, IL 61801-3003, USA
\\
$^3$Theoretical Physics Department, CERN, 1211 Geneva 23, Switzerland
}

\preprint{CERN-TH-2024-012}
\begin{abstract}
We show that it is possible to define a timelike future-directed information current within relativistic first-order hydrodynamics. This constitutes the first step towards a covariantly stable and causal formulation of first-order fluctuating hydrodynamics based on thermodynamic principles. We provide several explicit examples of first-order theories with an information current, covering many physical phenomena, ranging from electric conduction to viscosity and elasticity. We use these information currents to compute the corresponding equal-time correlation functions, and we find that the physically relevant (equal-time) correlators do not depend on the choice of the hydrodynamic frame as long as the frame leads to causal and stable dynamics. In the example of chiral hydrodynamics, we find that circularly polarized shear waves have different probabilities of being excited depending on their handedness, generating net helicity in chiral fluids.
\end{abstract}

\maketitle

\section{Introduction}

Understanding the subtle interplay between infrared (IR) and ultraviolet (UV) phenomena is one of the most serious challenges of relativistic fluid dynamics \cite{Kost2000,GavassinoFronntiers2021,GavassinoSuperluminal2021}. Recently, a systematic investigation into this issue was pursued in \cite{Bemfica2017TheFirst,BemficaDNDefinitivo2020,GavassinoGibbs2021,GavassinoUniversality2023} starting from fundamental physical principles. In a nutshell, the problem is that even if hydrodynamics is supposed to be applicable in the limit of very small gradients (the IR regime) \cite{Baier:2007ix,Romatschke:2017ejr,Kovtun2019}, still, a ``strict IR limit'' can never be achieved in finite systems because the Fourier transform of a compactly supported profile must contain all wavenumbers \cite{Hormander_book}. Hence, high-frequency (i.e., UV) contamination is unavoidable in practice, which becomes especially relevant in the nonlinear regime. In fact, UV issues can spoil both the physical content and the mathematical consistency of the equations, making some hydrodynamical theories essentially unphysical and ill-suited for applications \cite{Hiscock_Insatibility_first_order,Speranza2021,OlsonRegular1990,Bobylev1982}. 

The simplest example of a UV-pathological model is the linearized Super-Burnett diffusion equation \cite{Bobylev1982,SHAVALIYEV1993,Chernov2000,Struchup2005},
\begin{equation}\label{Burnett}
    \partial_t n = D \partial^2_x n +B \partial^4_x n\, ,
\end{equation}
where $n$ is a conserved density, $D{>}0$ is the Fick diffusion coefficient, and $B{>}0$ is the Super-Burnett coefficient\footnote{While the second law of thermodynamics forces $D$ to be positive, the sign of $B$ cannot be argued from such universal arguments. However, this coefficient turns out to be positive in kinetic theory models \cite{Struchup2005}.}. For small gradients, \eqref{Burnett} seems reasonable, as it implements the lowest-order correction to Fick's law. However, in the UV limit (i.e., outside the regime of validity of the theory where $k{\rightarrow} {+}\infty$, with $k$ being the wavenumber), the growth rate of the Fourier modes, $\Gamma(k){=}{-}Dk^2{+}Bk^4$, is large and positive. This results in a violent UV instability, which causes the initial value problem of \eqref{Burnett} to be ill-posed \cite{Bobylev1982}. Indeed, there are no solutions of equation \eqref{Burnett} for most initial data (including all Gaussian profiles).
To fix this kind of problem, one needs to regularize the UV, i.e. to introduce some additional term that acts to suppress high frequencies while being negligible at low frequencies. For example, in equation \eqref{Burnett}, we may add a fictitious term $\Lambda\partial^6_x n$ (with $\Lambda{>}0$) on the right side of \eqref{Burnett}, so that now $\Gamma(k){=}{-}Dk^2{+}Bk^4{-}\Lambda k^6$. This causes all the wavenumbers above the cutoff scale $k_c {\sim} \sqrt{B/\Lambda}$ to be automatically suppressed, making the initial value problem solvable (while leaving the IR unaffected).

Recently, a similar strategy has been adopted to regularize relativistic first-order hydrodynamic theories: the so-called Bemfica-Disconzi-Noronha-Kovtun (BDNK) approach \cite{Bemfica2017TheFirst,Kovtun2019,Bemfica2019_conformal1,Hoult:2020eho,BemficaDNDefinitivo2020} (with the most general form including both vector and axial-vector currents studied in \cite{Abboud:2023hos}). This includes additional first-order derivative corrections to the constitutive relations, which do not appear in the standard ``Navier-Stokes formulations'' of \citet{Eckart40} and \citet{landau6}. As it turns out, these additional terms do not modify the IR behavior because there is always a field redefinition that maps BDNK into standard Navier-Stokes plus some higher-order corrections \cite{Kovtun2019}. However, with an appropriate choice of parameters, these additional pieces can keep the UV sector under control, making BDNK (covariantly \cite{GavassinoBounds2023}) stable, causal, and its initial value problem well-posed \cite{BemficaDNDefinitivo2020}.

Paradoxically, while this technical improvement made it possible to rigorously solve relativistic first-order hydrodynamics numerically in complex situations (see, for example, \cite{Pandya:2021ief,Pandya:2022pif,Pandya:2022sff}), it also made it more difficult to implement stochastic fluctuations within a first-order framework \cite{Abbasi:2022rum,Mullins:2023ott,Jain:2023obu}. In fact, it has been recently shown that while the additional ``non-Navier-Stokes'' terms in BDNK regularize the deterministic equations of motion, they also add divergent UV contributions to the noise, causing fluctuations to grow seemingly out of control \cite{Mullins:2023ott} (see also  \cite{Jain:2023obu}). To understand why this happens, one should remember that BDNK hydrodynamics achieves well-posedness and covariant stability at the expense of not exactly satisfying the maximum entropy principle and the second law of thermodynamics in the UV \cite{GavassinoLyapunov_2020,DoreGavassino2022}. In fact, in BDNK hydrodynamics, these properties are only valid within the regime of validity of this effective theory. Given that the entropy determines the probability distribution of fluctuations, it is unsurprising that there can be a spontaneous condensation of all those UV modes whose entropy is higher than the equilibrium entropy in the current description of fluctuating first-order theories.

A way out is to regularize the probability distribution of fluctuations of first-order theories. In particular, one should be able to add some UV regulators to the entropy, making the latter \textit{exactly} non-decreasing along solutions of the corresponding first-order theory and \textit{exactly} maximal at equilibrium, also in the UV limit. In the linear regime, this corresponds to finding an information current $E^\mu$ and an entropy production rate $\sigma$, both quadratic in deviations from global equilibrium, with the following properties \cite{GavassinoCausality2021}:
\begin{itemize}
    \item[(i)] For fluctuations that conserve the values of all the integrals of motion of the system, we must have that \cite{GavassinoGibbs2021}
    \begin{equation}
        E=\int_{\mathbb{R}^3} E^0 \, d^3 x = S_{\text{eq}}-S \, ,
    \end{equation}
    at least up to first-order in gradients ($S=$ entropy of the state, $S_{\text{eq}}=$ entropy at equilibrium). This implies that if we know the constitutive relation for the entropy current to first order (see e.g. Appendix A of \cite{Kovtun2019}), then the UV regularized $E^\mu$ should agree to first order with the $E^\mu$ computed from the direct variation of the entropy. 
    \item[(ii)] The entropy production equation $\partial_\mu E^\mu+\sigma=0$ must hold as an exact identity along all classical solutions of the equations of motion of the first-order theory. Thus, one requires the second law of thermodynamics to hold even in the UV limit, i.e. far beyond the regime of applicability of the effective theory.   
    \item[(iii)] The vector $E^\mu$ should always be timelike future-directed\footnote{We note that $E^\mu$ is also allowed to be lightlike, future directed. However, it considerably simplifies the analysis if we work with strict inequalities and deal with limiting cases afterward, on a case-by-case basis.}, and the scalar $\sigma$ should always be non-negative, for arbitrary perturbations, both on-shell and off-shell. In other words, the maximum entropy principle should also hold in the UV limit and independently of whether the fluctuations are solutions to the equations of motion.
\end{itemize}
Under these assumptions, one can not only automatically prove stability and causality of the linearized equations of motion \cite{GavassinoGibbs2021,GavassinoCausality2021} but also of Gaussian stochastic fluctuations because the probability distribution $e^{S}$ is now maximized in equilibrium. Furthermore, one is also able to set up a fully covariant theory of fluctuations directly from the knowledge of $E^\mu$ and $\sigma$ \cite{Mullins:2023tjg}, through a relativistic generalization of the Fox-Uhlenbeck approach \cite{FoxUhlFluctuations1970}.

In this work, we construct a UV-regularized information current and entropy production rate for many first-order theories, so that requirements (i,ii,iii) are all fulfilled. Unfortunately, the procedure often involves some trial and error, and it is unclear at this point that a general systematic derivation exists. However, we will provide a long list of concrete examples pertaining to deeply different physical systems so that the reader should have an overview of the typical information currents that arise in this procedure. As a first application in the direction of fluctuating hydrodynamics, we will compute all the equal-time correlators of the associated theories. The development of a consistent first-order theory of stochastic fluctuations using the insights obtained in this work will be carried out in another paper \cite{GavassinoFluctatingBDNK2024vyu}. 

This paper is organized as follows. In Section \ref{doodlebob}, we explain why standard methods lead to an inadequate information current in first-order theories, and we outline a regularization procedure that can be used to fix the problem on a case-by-case basis. In Section \ref{bazinga!}, we use this procedure to construct a regularized information current for the relativistic diffusion equation, and we illustrate its utility by computing equal-time correlators in the fluctuating theory. The remaining sections follow more or less the same format as this, showcasing the method with examples from different physical contexts. In Section \ref{Electro!}, we consider a diffusing charge coupled to electric and magnetic fields. In Sections \ref{bulkuzzo} and \ref{shearuzzuzuzuz}, we consider simple causal and stable first-order theories for relativistic fluids with bulk viscosity (Section \ref{bulkuzzo}) and shear viscosity (Section \ref{shearuzzuzuzuz}). In Section \ref{tinytim}, we depart from dissipative models, turning our attention to the non-dissipative dynamics of Goldstone modes. We consider Goldstone modes associated with a spontaneously broken $U(1)$ symmetry (as in a superfluid) or translational symmetry (as in an elastic medium). In Section \ref{CHIRUZ}, we consider ideal chiral hydrodynamics, the non-dissipative theory of a relativistic fluid with an underlying chiral anomaly. We conclude in Section \ref{conk} with some physical remarks. Throughout the article, we adopt the spacetime signature $(-,+,+,+)$ in Minkowski spacetime with Cartesian coordinates and natural units $c=k_B=\hbar=1$. Greek indices are spacetime indices that run from 0 to 3, while Latin indices are purely spatial indices running from 1 to 3 (Einstein's convention applies to all indices). For the volume form, we adopt the convention $\varepsilon_{0123}=+1$.

\section{UV-regularized information current} \label{doodlebob}

Previous works argued that first-order hydrodynamics cannot admit a proper information current \cite{GavassinoLyapunov_2020,DoreGavassino2022}. This is due to the fact that quadratic vector fields that are truncated to first order in derivatives always fail to be timelike future directed at large gradients. Here, we discuss the origin of the problem and present our solution.


\subsection{First-order truncation}

Let $\varphi$ be the collection of linearized fields describing small infrared perturbations in a fluid. By definition, $\varphi$ vanishes in global equilibrium. Assuming that the background equilibrium state is homogeneous and isotropic and the spacetime is flat, the information current and entropy production rate should be expressed as local functions of the perturbation fields and their derivatives, namely
\begin{equation}\label{regina}
\begin{split}
E^\mu ={}& E^\mu (\varphi, \partial_\nu \varphi, \partial_\nu \partial_\rho \varphi,...) \, , \\
\sigma ={}& \sigma (\varphi, \partial_\nu \varphi, \partial_\nu \partial_\rho \varphi,...) \, . \\
\end{split}
\end{equation} 
Given that $E^\mu$ and $\sigma$ are constructed as second variations \cite{GavassinoGibbs2021}, they are quadratic in the fluctuations. Hence, they are linear combinations of terms of the form $\partial^m \varphi \, \partial^n \varphi$, where $m$ and $n$ denote the number of derivatives. In a hypothetical ``exact theory'', \eqref{regina} may contain an infinite series of terms.  However, since the hydrodynamic theory applies only to small gradients, one can perform a truncation at some finite order in derivatives. In practice, if we decide to truncate at order $r$, then we only need to include all terms $\partial^m \varphi \, \partial^n \varphi$ with $m+n\leq r$. 

If we truncate to zeroth order, there are no derivatives, and we obtain the information current in the ideal fluid limit. Thus, if we want to include viscous corrections, we must keep at least the first-order terms. If we decide to truncate only up to the first order, then we can schematically write (to lighten the notation, the linear combination coefficients are understood) 
\begin{equation}
E^\mu \sim \varphi \varphi + \varphi \partial \varphi \, .
\end{equation}
Now we immediately find an issue. In fact, we would like $E^\mu$ to be \textit{always} timelike future directed. Hence, we should require $E^0-E^1$ to be a positive definite quadratic form. But, in general, we have
\begin{equation}
E^0-E^1 \sim \varphi \varphi + \varphi \partial \varphi  \sim (\varphi, \partial \varphi)
\begin{bmatrix}
\#  &  \# \\
\# &  0 \\
\end{bmatrix}
\begin{pmatrix}
\varphi \\
\partial \varphi \\
\end{pmatrix} \, .
\end{equation}
Clearly, if the $\varphi \partial \varphi$ term does not vanish, the quantity $E^0-E^1$ cannot be a non-negative definite quadratic form due to the zero on the diagonal. Hence, $E^\mu$ fails to be a timelike future-directed 4-vector at large gradients. 

From a purely physical perspective, this does not contradict any thermodynamic principle because only the infinite series \eqref{regina} must truly be timelike future-directed. Indeed, it is evident that $E^0-E^1$ becomes negative only when the term $\varphi \partial \varphi$ becomes comparable to the term $\varphi \varphi$, which is precisely when the first order truncation is no longer applicable (thus, outside of the regime of validity of the first-order theory). However, if the truncated $E^\mu$ is not timelike future directed in the UV, then the corresponding probability distribution $e^{S-S_{\text{eq}}}=e^{-E}$ is ill-defined due to unphysical UV divergences\footnote{\label{This_orsiu}The proof is straightforward. Suppose that some fluctuation $\bar{\varphi}$ is such that $\bar{E}<0$. Then, the perturbations $a\bar{\varphi}$ (with $a =\text{const}$) have probability $\propto e^{-a^2\bar{E}}$, which tends to $+\infty$ for large $a$. The result is a non-normalizable (and therefore ill-defined) probability distribution, which favors infinitely large deviations from equilibrium. On the other hand, if $\bar{E}$ is positive, then $e^{-a^2\bar{E}}$ is a Gaussian peaked at $a=0$.}. This makes the truncated information current useless for studying fluctuations. Hence, our goal now is to introduce some higher-frequency corrections, which regularize the UV sector by introducing a cutoff, thereby suppressing all unphysical divergences. 

\subsection{UV regularization}

From a physical standpoint, there is no reason to expect that $E^\mu$, when truncated at \textit{any} given (non-zero) order, remains timelike future directed also at large gradients. Indeed, even if we go up to second order, by including terms of the form $\partial \varphi \partial \varphi$ and $\varphi \partial \partial \varphi$, we have
\begin{equation}
E^0-E^1 \sim  (\varphi, \partial \varphi, \partial \partial \varphi)
\begin{bmatrix}
\#  &  \# & \# \\
\# &  \# & 0 \\
\# &  0  &  0 \\
\end{bmatrix}
\begin{pmatrix}
\varphi \\
\partial \varphi \\
\partial \partial \varphi\\
\end{pmatrix} \, ,
\end{equation}
and, again, there is a $0$ on the diagonal. 

Luckily, there is a simple solution. Let us recall that the entropy current of first-order theories is determined only up to first order in derivatives \cite{Kovtun2019}. Hence, the terms $\partial \varphi \partial \varphi$ and $\varphi \partial \partial \varphi$ are of the same order as the truncation error, and their exact value is irrelevant for physical purposes. This implies that we can ``fix them at will'' in this order. In particular, one can set the terms $\varphi \partial \partial \varphi$ to zero, so that now
\begin{equation}\label{five}
E^0-E^1 \sim  (\varphi, \partial \varphi)
\begin{bmatrix}
\#  &  \# \\
\# &  \# \\
\end{bmatrix}
\begin{pmatrix}
\varphi \\
\partial \varphi \\
\end{pmatrix} \, .
\end{equation}
Then, we just need to tune the terms $\partial \varphi \partial \varphi$ to make $E^\mu$ timelike future directed also at large gradients. At the same time, we can try to enforce that the equation $\partial_\mu E^\mu +\sigma=0$ be \textit{exactly} consistent with the equations of motion (also outside the formal regime of applicability of the theory). When this is possible, we can say that a first-order theory admits a proper (i.e. useful) information current, and it can therefore be made stable also against stochastic fluctuations \cite{Mullins:2023tjg}. Note that, in this setting, the terms $\partial \varphi \partial \varphi$ should always be interpreted as mere UV regulator terms, suppressing spurious UV cutoff phenomena in the fluctuating system. For this reason, the detailed structure of these second-order terms is highly sensitive to the behavior of the theory at large gradients. This implies that the terms $\partial \varphi \partial \varphi$ cannot have simple (and universal) transformation laws under changes of hydrodynamic frame, since the latter abruptly redefine both the causality and the stability properties of the UV sector.

\subsection{Quick application: equal-time correlators}

The equilibrium probability distribution of fluctuations of an isolated system is proportional to $e^{S-S_{\text{eq}}}=e^{-E}$. Hence, one can use the information current to compute any equal-time correlation function by averaging over all possible hydrodynamic macrostates at a given time. The result is a functional integral, e.g.
\begin{equation}\label{functionalInt}
    \langle \varphi(\textbf{x}) \varphi^T(\textbf{y}) \rangle = \dfrac{\int  \varphi(\textbf{x}) \varphi^T(\textbf{y}) e^{-E} \, \mathcal{D} [\varphi] \, \mathcal{D}[\partial_t \varphi]}{\int  e^{-E} \, \mathcal{D} [\varphi] \, \mathcal{D}[\partial_t \varphi] } \, ,
    \end{equation}
which converges in a fully covariant manner provided that $E^\mu$ is timelike future directed (see footnote \ref{This_orsiu}). 
Note that the functional integral must be performed over all the field configurations on a three-dimensional $t=\text{const}$ hypersurface (not on the whole 4D spacetime). Hence, if the equation of motion is of second order in time, the time derivatives $\partial_t \varphi$ must be treated as independent degrees of freedom since they can be chosen freely, and they contribute to defining the physical state at $t=0$. For this reason, we needed to include them in the measure of \eqref{functionalInt} as additional variables.

Since the information current is quadratic in the fields, equation \eqref{functionalInt} defines a Gaussian functional integral, which can be evaluated using standard field theory techniques \cite{weinbergQFT_1995}. In fact, if we rewrite $E$ as a generalized quadratic form,
\begin{equation}\label{mazurca}
    E= \int d^3x \,  \dfrac{1}{2} (\varphi^T, \partial_t \varphi^T)
\mathcal{K}(\partial_j) \!
\begin{pmatrix}
\varphi \\
\partial_t \varphi \\
\end{pmatrix} 
=
\int d^3x \, d^3 y \, \dfrac{1}{2} (\varphi^T(\textbf{x}), \partial_t \varphi^T(\textbf{x}))
\mathcal{K}(\textbf{x}-\textbf{y}) \!
\begin{pmatrix}
\varphi(\textbf{y}) \\
\partial_t \varphi(\textbf{y}) \\
\end{pmatrix} \, , 
\end{equation}
where $\mathcal{K}$ is Hermitian, and the kernel is
\begin{equation}
    \mathcal{K}(\textbf{x}-\textbf{y}) = \int \dfrac{d^3 k}{(2\pi)^3} e^{i\textbf{k}\cdot (\textbf{x}-\textbf{y})} \mathcal{K}(ik_j) \, ,
\end{equation}
then we have the following well-established formula for the correlators \cite{weinbergQFT_1995} ($\mathcal{K}^{-1}$ is the ordinary matrix inverse of $\mathcal{K}$):
\begin{equation}\label{feynman}
\begin{bmatrix}
  \langle \varphi(\textbf{x}) \, \varphi^T(\textbf{y}) \rangle   &  \langle \varphi(\textbf{x}) \, \partial_t \varphi^T(\textbf{y}) \rangle\\
   \langle \partial_t \varphi(\textbf{x}) \, \varphi^T(\textbf{y}) \rangle  & \langle \partial_t  \varphi(\textbf{x}) \, \partial_t \varphi^T(\textbf{y}) \rangle \\
\end{bmatrix}
     = \int \dfrac{d^3 k}{(2\pi)^3} e^{i\textbf{k}\cdot (\textbf{x}-\textbf{y})} \mathcal{K}(ik_j)^{-1} \, .
\end{equation}
It is also possible to use the information current to evaluate correlators at non-equal times. This task requires the construction of a fully dynamic theory of stochastic fluctuations mirroring \cite{Mullins:2023tjg}, which we leave for future work \cite{GavassinoFluctatingBDNK2024vyu}.

\section{Causal diffusion}\label{bazinga!}

We begin with the simplest example possible: the relativistic diffusion equation. Let $\varphi$ be the perturbation to the baryon chemical potential (possibly rescaled by some background constant). Assume that, in the equilibrium global rest frame of the medium, $\varphi$ obeys the equation
\begin{equation}\label{Cattaneo}
\tau \partial^2_t \varphi +\partial_t \varphi -D \partial_j \partial^j \varphi=0 \, .
\end{equation}
The transport coefficients $\tau$ and $D$ are positive constants, and causality demands that $\tau>D$. Equation \eqref{Cattaneo} can be viewed as a first-order theory\footnote{Equation \eqref{Cattaneo} can also be interpreted as an Israel-Stewart theory \cite{Israel_Stewart_1979,cattaneo1958}, in which case we would need to introduce and independent degree of freedom $q^j$, satisfying the equations $\partial_t \varphi +\partial_j q^j=0$ and $\tau \partial_t q_j +q_j =-D\partial_j \varphi$, and the information current would be a function of both $\varphi$ and $q^j$ \cite{GavassinoNonHydro2022,GavassinoUniversality2023}.}, with a conserved baryon four-current $J^\mu =(\varphi+\tau \partial_t \varphi, -D \partial^j \varphi)$. Note that the value of $\tau$ defines the choice of the hydrodynamic frame since we can always make a field redefinition $\varphi \rightarrow \varphi + a \partial_t \varphi$ \cite{Kovtun2019}.

\subsection{Regularized information current}

It is easy to verify that, up to a global multiplicative constant, the most general information current and entropy production rate for this model, involving only terms as in \eqref{five}, are
\begin{equation}\label{INFOONA}
\begin{split}
E^0 ={}& \dfrac{1}{2} \big[ \varphi^2 + 2 \tau \varphi \partial_t \varphi + \lambda \tau (\partial_t \varphi)^2+ \lambda D \partial_j \varphi \partial^j \varphi \big] \, ,\\
E^j ={}& -D(\varphi + \lambda \partial_t \varphi)\partial^j \varphi \, ,\\
\sigma ={}& (\lambda-\tau)(\partial_t \varphi)^2 + D \partial_j \varphi \partial^j \varphi \, , \\
\end{split}
\end{equation}
where $\lambda$ is an additional free parameter (to be fixed to ensure thermodynamic stability \cite{GavassinoCausality2021}). We note that \eqref{INFOONA} contains all the terms allowed by symmetry, apart from the terms $\varphi^2$ and $\varphi \partial_t \varphi$ in the entropy production rate, which are forbidden by baryon conservation and the second law of thermodynamics in the infrared limit. Furthermore, the coefficients have been fixed in such a way that the equation $\partial_\mu E^\mu +\sigma=0$ takes the \textit{exact} form
\begin{equation}
(\varphi+\lambda \partial_t \varphi)(\tau \partial^2_t \varphi +\partial_t \varphi-D\partial_j \partial^j \varphi)=0 \, ,
\end{equation}
which is automatically satisfied if the equation of motion \eqref{Cattaneo} holds. Finally, thermodynamic stability holds provided that $\lambda>\tau>D>0$. In fact, $\sigma$ is positive definite if and only if $\lambda > \tau$ and $D >0$. Furthermore, the quantity $E^0-E^1$ can be expressed as follows:
\begin{equation}\label{gargarensis}
E^0-E^1 = \dfrac{1}{2} \big(\varphi, \partial_t \varphi, \partial_1 \varphi \big)
\begin{bmatrix}
1 & \tau & D \\
\tau & \lambda \tau & \lambda D \\
D & \lambda D & \lambda D \\
\end{bmatrix}
\begin{pmatrix}
\varphi \\
\partial_t \varphi \\
\partial_1 \varphi \\
\end{pmatrix}
+ \dfrac{1}{2} \lambda D \big[ (\partial_2 \varphi)^2+(\partial_3 \varphi)^2 \big]
\end{equation}
The positivity of the diagonal elements of the $3 \times 3 $ matrix implies $\lambda>0$ and $\tau>0$. The  determinants of the $2\times 2$ blocks are $\tau(\lambda-\tau)$, $\lambda^2 D(\tau-D)$, and $D(\lambda-D)$, and their positivity implies $\lambda>\tau$, $\tau>D$, and $\lambda>D$. Finally, the determinant of the $3 \times 3$ matrix itself is $D\lambda (\lambda-\tau)(\tau-D)$, which is also positive under the above conditions. Collecting together all the inequalities above, one obtains $\lambda>\tau>D>0$.

It should be noted that $\tau>D>0$ is indeed the condition for covariant stability of the field equation \eqref{Cattaneo}. On the other hand, we note that $\lambda$ does not appear in the equation of motion. Instead, it is necessary for the system's stability against stochastic fluctuations (i.e., off-shell). In fact, the probability distribution of small fluctuations is proportional to $e^{-E}$, where $E=\int E^0 d^3 x$ is given by
\begin{equation}\label{leilei}
E= \dfrac{1}{2} \int  \bigg[ (\varphi+\tau \partial_t \varphi)^2 + \tau(\lambda -\tau) (\partial_t \varphi)^2+ \lambda D \partial_j \varphi \partial^j \varphi \bigg] d^3 x \, .
\end{equation} 
One can see that the condition $\lambda>\tau$ is required to guarantee that all perturbations have a positive-definite free energy cost and are, therefore, less probable than the equilibrium state.

Let us now comment on the limiting case $\lambda =\tau$. Earlier, this case was automatically excluded because it makes the quadratic form \eqref{leilei} only non-negative definite instead of strictly positive. However, we see from \eqref{leilei} that the only fluctuation with vanishing $E$ is the non-hydrodynamic mode $\varphi=e^{-t/\tau}$, which is a set of measure zero in the state-space. Hence, it does not lead to any pathologies in the probability distribution $e^{-E}$. Actually, it is reasonable that this non-hydrodynamic mode does not affect the entropy, as it is a pure ``frame relaxation'', which has no impact on the conserved fluxes and is a mathematical artifact of the theory.  Indeed, for $\lambda=\tau$, equation \eqref{leilei} becomes
\begin{equation}\label{leilei2}
E= \dfrac{1}{2} \int  \bigg[ (J^0)^2 +  \dfrac{\tau}{D} J^j J_j \bigg] d^3 x \, ,
\end{equation} 
which depends only on the physical flux $J^\mu=(\varphi+\tau \partial_t \varphi,-D\partial^j \varphi)$, and not on the value of $\varphi$ itself. Below, we will see that $\lambda=\tau$ is actually the most convenient (and physically appealing) option in a theory of fluctuations.

\subsection{Equal time correlators}\label{causuzzuz}

The UV-regularized information current can be used to define a probability distribution $e^{-E}$ for the stochastic fluctuations, and we can use it to compute equal time correlators. In  fact, if we integrate the last term by parts,  the total information \eqref{leilei} can be expressed in the form \eqref{mazurca}, with
\begin{equation}
    \mathcal{K}(\partial_j) = 
    \begin{bmatrix}
     1 {-}\lambda D \partial^j \partial_j &  & \tau \\
     \tau &  & \lambda \tau \\
    \end{bmatrix} \, .
\end{equation}
Then, equation \eqref{feynman} becomes\footnote{In equation \eqref{INFOONA}, the information current had been rescaled for convenience. Consequently, there should be an overall constant multiplying the correlators \eqref{feynman2}. For simplicity, we work in some units where such a constant is still $1$.}
\begin{equation}\label{feynman2}
\begin{bmatrix}
  \langle \varphi(\textbf{x}) \, \varphi(\textbf{y}) \rangle   &  \langle \varphi(\textbf{x}) \, \partial_t \varphi(\textbf{y}) \rangle\\
   \langle \partial_t \varphi(\textbf{x}) \, \varphi(\textbf{y}) \rangle  & \langle \partial_t  \varphi(\textbf{x}) \, \partial_t \varphi(\textbf{y}) \rangle \\
\end{bmatrix}
     = \int \dfrac{d^3 k}{(2\pi)^3} \, \dfrac{ e^{i\textbf{k}\cdot (\textbf{x}-\textbf{y})}}{(1{+}\lambda D k^j k_j)\lambda \tau {-}\tau^2} \, \begin{bmatrix}
         \lambda \tau & -\tau \\
         -\tau &  1{+}\lambda D k^j k_j \\
     \end{bmatrix} \, .
\end{equation}
These integrals can be evaluated analytically. For example, the field-field correlator obeys a Yukawa-type decay law:
\begin{equation}\label{Yuks}
   \langle \varphi(\textbf{x}) \, \varphi(\textbf{y}) \rangle  = \dfrac{e^{-a|\textbf{x}-\textbf{y}|}}{4\pi \lambda D |\textbf{x}-\textbf{y}|} \, , \spc \text{with }a=\sqrt{\dfrac{\lambda-\tau}{\lambda^2 D}} \, . 
\end{equation}
However, one should remember that the field $\varphi$ does not have a particularly interesting physical meaning because it is related to a specific choice of hydrodynamic frame. Instead, we should focus on correlators involving, e.g., $J^0 {=} \varphi {+}\tau \partial_t \varphi$, which is an unambiguously defined conserved density. Interestingly, we find that, if we set $\lambda{=}\tau$, then the density-density correlator reduces to $\langle J^0(\textbf{x})  J^0(\textbf{y}) \rangle= \delta^3(\textbf{x}{-}\textbf{y})$. This is a useful result because it does not depend on $\tau$ (i.e. on our choice of hydrodynamic frame). Furthermore, this formula is fully consistent with its non-relativistic analog; see section \S 88 of \cite{landau9}.

\section{Causal electric conduction}\label{Electro!}

Let us now consider an extension of the previous model, where a coupling with an electromagnetic field is included.

\subsection{Overview of the model}

We consider a linear theory for three fields, $\{\varphi, \mathcal{E}_j,\mathcal{B}_j \}$, representing respectively the perturbation to the charge chemical potential, the electric field, and the magnetic field (possibly rescaled by some background constants). We assume, for simplicity, that the medium is an isotropic conductor, with conductivity $\Sigma$, so that the electric current takes the form $J^\mu=(\varphi+\tau \partial_t \varphi,-D \partial^j \varphi + \Sigma \mathcal{E}^j)$, where the term $\Sigma \mathcal{E}^j$ is the drift flux due to Ohm's law. Then we can write Maxwell's equations in a conductor (for simplicity, we ignore polarization and magnetization corrections):
\begin{equation}\label{maxone}
\begin{split}
& \partial_j \mathcal{E}^j = \varphi +\tau \partial_t \varphi \, , \\
& \partial_j \mathcal{B}^j = 0 \, , \\
& (\nabla \times \mathcal{E})_j =-\partial_t \mathcal{B}_j \, , \\
& (\nabla \times \mathcal{B})_j =-D \partial_j \varphi +\Sigma \mathcal{E}_j +\partial_t \mathcal{E}_j \, . \\
\end{split}
\end{equation}
The first equation can be viewed as an equation of motion for $\varphi$, the third as an equation of motion for $\mathcal{B}_j$, and the fourth as an equation of motion for $\mathcal{E}_j$. Hence, they are enough to fully determine the evolution of the system. The second equation, instead, represents a constraint on the initial data. In fact, if we take the divergence of the third equation, we obtain $\partial_t (\partial_j \mathcal{B}^j)=0$, which automatically implies $\partial_j \mathcal{B}^j =0$, provided that the latter equation holds already in the initial state. Note that if we take the divergence of the fourth equation and use the first, we obtain the law of charge conservation $\partial_\mu J^\mu =0$, namely
\begin{equation}\label{ioconservoetu?}
\partial_t \varphi +\tau \partial^2_t \varphi -D \partial_j \partial^j \varphi + \Sigma \partial_j \mathcal{E}^j =0 \, .
\end{equation}

Let us now derive the conditions that ensure stability and causality for this system. This is easiest to do if we rewrite the dynamical part of the system \eqref{maxone} in the following form:
\begin{equation}\label{gargantua}
\begin{split}
& \tau \partial^2_t \varphi +(1+\tau \Sigma )\partial_t \varphi +\Sigma \varphi -D \partial_j \partial^j \varphi =0  \, ,\\
& \partial^2_t \mathcal{B}_j +\Sigma \partial_t \mathcal{B}_j-\partial_k \partial^k \mathcal{B}_j =0  \, ,\\
& \partial_t \mathcal{E}_j + \Sigma \mathcal{E}_j =(\nabla \times \mathcal{B})_j +D \partial_j \varphi  \, .\\
\end{split}
\end{equation}
As can be seen, the fields $\varphi$ and $\mathcal{B}_j$ obey decoupled evolution equations. The first represents the tendency of charge density to diffuse and neutralize, while the second is the equation of magnetic diffusion. It is straightforward to verify that, for the first two equations to be causal and stable, we only need to set $\tau >D>0$ and $\Sigma>0$. To prove that also the evolution of the electric field is stable, it is enough to notice that the third equation of \eqref{gargantua} is a relaxation equation (since $\Sigma>0$) with a source. Given that the source depends linearly on $\varphi$ and $\mathcal{B}_j$, which decay to zero at late times (by diffusion),  $\mathcal{E}_j$ must also relax to zero, proving that the system is indeed stable. Let us finally show that the electric field propagates subluminally. To this end,  assume that $\varphi$, $\mathcal{E}_j$ and $\mathcal{B}_j$ (and thus also $\partial_t \varphi \propto \partial_j \mathcal{E}^j-\varphi$, and $\partial_t \mathcal{B} \propto \nabla {\times} \mathcal{E}$) are initially supported inside a compact set $\mathcal{R}$. Then we know from \eqref{gargantua} that $\varphi$ and $\mathcal{B}_j$ propagate inside the future lightcone of $\mathcal{R}$. Hence, outside of such lightcone, the electric field is a solution of the equation $\partial_t \mathcal{E}_j + \Sigma \mathcal{E}_j=0$, with initial data $\mathcal{E}_j(t{=}0)=0$. The only solution is $\mathcal{E}_j=0$, proving that the system is indeed causal.

\subsection{Information current and entropy production rate}

We managed to find, with some trial and error, two (non-equivalent) alternative formulas for the information current and the associated entropy production rate (both rescaled by an overall constant). Below, we present the simpler construction, while the other construction is reported in Appendix \ref{AAA}.
\begin{equation}\label{INFOONAelectro}
\begin{split}
E^0 ={}& \dfrac{1}{2} \bigg[ (\varphi +  \tau  \partial_t \varphi)^2 + \tau D \partial_j \varphi \partial^j \varphi +\dfrac{\Sigma}{D} (\mathcal{E}^j \mathcal{E}_j + \mathcal{B}^j \mathcal{B}_j) \bigg] \, ,\\
E^j ={}& -D(\varphi + \tau \partial_t \varphi)\partial^j \varphi +\dfrac{\Sigma}{D} (\mathcal{E} \times \mathcal{B})^j \, ,\\
\sigma ={}&  D \partial_j \varphi \partial^j \varphi-\Sigma \mathcal{E}^j \partial_j \varphi+\dfrac{\Sigma^2}{D}\mathcal{E}^j \mathcal{E}_j+ \Sigma (\varphi+\tau \partial_t \varphi)^2 \, . \\
\end{split}
\end{equation}
To see that \eqref{INFOONAelectro} is \textit{exactly} consistent with the Maxwell equations, we note that the condition $\partial_\mu E^\mu +\sigma=0$ takes the explicit form
\begin{equation}
\begin{split}
& (\varphi+\tau \partial_t \varphi)(\partial_t \varphi +\tau \partial^2_t \varphi - D \partial_j \partial^j \varphi + \Sigma \partial_j \mathcal{E}^j) + \Sigma (\varphi +\tau \partial_t \varphi)(\varphi +\tau \partial_t \varphi-\partial_j \mathcal{E}^j)\\
& +\dfrac{\Sigma}{D} \mathcal{E}^j \big[ \partial_t \mathcal{E}_j-D\partial_j \varphi+\Sigma \mathcal{E}_j-(\nabla \times \mathcal{B})_j \big]+\dfrac{\Sigma}{D}\mathcal{B}^j \big[ \partial_t \mathcal{B}_j +(\nabla \times \mathcal{E})_j \big]=0 \, , \\
\end{split}
\end{equation}
which, indeed, vanishes identically along all exact solutions of the equations of motion \eqref{maxone}.

Let us comment on the expressions in \eqref{INFOONAelectro}. First, note that the electromagnetic contributions to the information current have the prefactor $\Sigma/D$, and the contribution to $E^0$ is the free energy of the electromagnetic field. This leads us to the Wiedemann–Franz law, which states that the quotient $\Sigma/D$ is a purely thermodynamic (i.e. non-kinetic) parameter that can be computed directly from the equation of state \cite{GavassinoMaxwell2023}. Secondly, we note that, in the infrared limit, the formula for $\sigma$ is consistent with the Israel-Stewart dissipation equation $\sigma=J_j J^j/D$ \cite{GavassinoMaxwell2023}. In fact, if we use \eqref{maxone} to replace one of the factors $\varphi +\tau \partial_t \varphi$ with $\partial_j \mathcal{E}^j$ in the last term, we obtain 
\begin{equation}
\sigma = \dfrac{J^j J_j}{D} +\partial_j \big[ \Sigma J^0 \mathcal{E}^j \big]-\Sigma \tau \mathcal{E}^j \partial_t \partial_j \varphi \, .
\end{equation}
The first term on the right-hand side is indeed the Israel-Stewart dissipation rate. The second term is a pure divergence, whose contribution to the total entropy production vanishes when integrated over the whole space. The last piece is a third-order term (in conductors, $\mathcal{E}^j$ is considered of order 1 in derivatives \cite{Hernandez:2017mch}), which is negligible in the IR limit.

\subsection{Thermodynamic stability}

Let us now find the conditions under which $E^\mu$ is timelike future directed and $\sigma$ is non-negative. It is immediately evident that we must have $D>0$, $\tau>0$, and $\Sigma>0$. Furthermore, we have the following equations:
\begin{equation}
\begin{split}
2(E^0-E^1) ={}&  \big( J^0, \partial_1 \varphi \big)
\begin{bmatrix}
1 & D \\
D  & \tau D \\
\end{bmatrix}
\begin{pmatrix}
J^0 \\
\partial_1 \varphi \\
\end{pmatrix} 
+  \tau D \big[ (\partial_2 \varphi)^2 {+} (\partial_3 \varphi)^2 \big] + \dfrac{\Sigma}{D} \big[ \mathcal{E}^j \mathcal{E}_j + \mathcal{B}^j \mathcal{B}_j -2(\mathcal{E} {\times} \mathcal{B})^1 \big] \, ,\\
\sigma ={}& \big(\partial_j \varphi, \mathcal{E}_j \big) \begin{bmatrix}
D & -\Sigma/2 \\
-\Sigma/2 & \Sigma^2/D \\
\end{bmatrix}
\begin{pmatrix}
\partial^j \varphi \\ 
\mathcal{E}^j \\
\end{pmatrix}+ \Sigma (\varphi+\tau \partial_t \varphi)^2  \, . \\
\end{split}
\end{equation}
The determinant condition for the $2\times 2$ matrix in the first line produces the causality condition $\tau>D$. The electromagnetic contribution to $E^0-E^1$ is positive definite due to the well known inequality $||\mathcal{E} {\times} \mathcal{B} || \leq ||\mathcal{E}|| {\cdot} || \mathcal{B} || $. The $2 \times 2$ matrix on the second line is automatically positive definite, provided that $D>0$. Hence, we have recovered exactly the same causality and stability conditions of the equations of motion \eqref{maxone}.

\subsection{Equal time correlators}

In the functional integral, we need to sum over all the field configurations that are \textit{physically admissible}. This implies that we cannot include in the sum any arbitrary magnetic field configuration, but we need to guarantee that $\partial_j \mathcal{B}^j=0$ holds on all relevant states in the integral. The simplest way to enforce this is to add the term $\frac{1}{2}\Lambda (\partial_j \mathcal{B}^j)^2$ to $E^0$, and take the limit $\Lambda \rightarrow +\infty$ at the end of the calculation. This guarantees that all configurations with $\partial_j \mathcal{B}^j \neq 0$ have zero probability of occurring. Note that $\partial_\mu E^\mu +\sigma=0$ holds along solutions of \eqref{maxone} also with this new term. 

There is one more subtlety. In the case of diffusion without the electromagnetic field (Section \ref{causuzzuz}), we had to consider $\partial_t \varphi$ as an independent degree of freedom, because the equation of motion  \eqref{Cattaneo} is of second order in time. On the other hand, in the Maxwell system \eqref{maxone}, the first equation of motion fully determines the value of $\partial_t \varphi$ in terms of $\varphi$ and $\mathcal{E}^j$. Hence, we cannot treat $\partial_t \varphi$ as an independent degree of freedom. Instead, the free variables that define the measure in the functional integral are $\{\varphi,\mathcal{E}^j,\mathcal{B}^j\}$, since these are enough to fully determine the physical state at a given time. This implies that, in the formula for $E$, we need to replace $\partial_t \varphi$ with $(\partial_j \mathcal{E}^j-\varphi)/\tau$.

All of this leads us to the following formula:
\begin{equation}
    E = \dfrac{1}{2} \int \bigg[  \tau D \partial_j \varphi \partial^j \varphi +\dfrac{\Sigma}{D} \mathcal{E}^j \mathcal{E}_j +(\partial_j \mathcal{E}^j)^2 +\dfrac{\Sigma}{D} \mathcal{B}^j \mathcal{B}_j + \Lambda (\partial_j \mathcal{B}^j)^2 \bigg] d^3 x \, .
\end{equation}
As can be seen, there is no coupling between $\varphi$, $\mathcal{E}^j$, and $\mathcal{B}^j$, meaning that the cross-correlators $\varphi \mathcal{E}$, $\mathcal{E}\mathcal{B}$, and $\mathcal{B}\varphi$ vanish. Indeed, the vanishing of equal time $\mathcal{E}\mathcal{B}$ correlators is a well-known result of fluctuating electrodynamics \cite{landau9}. Furthermore, the $\varphi\varphi$ correlator is the same as \eqref{Yuks}, with $\lambda=\tau$. Hence, we can just focus on the $\mathcal{E}\mathcal{E}$ and $\mathcal{B}\mathcal{B}$ correlators, whose Fourier integrals are reported below (we have already sent $\Lambda \rightarrow +\infty$):
\begin{equation}
    \begin{split}
\langle \mathcal{E}_j(\textbf{x})\mathcal{E}_k(\textbf{y}) \rangle ={}& \dfrac{D}{\Sigma} \int \dfrac{d^3 k}{(2\pi)^3} e^{i\textbf{k}\cdot (\textbf{x}-\textbf{y})}  \bigg( \delta_{jk} - \dfrac{D k_j k_k}{\Sigma + Dk^l k_l} \bigg) \, , \\
\langle \mathcal{B}_j(\textbf{x})\mathcal{B}_k(\textbf{y}) \rangle ={}& \dfrac{D}{\Sigma} \int \dfrac{d^3 k}{(2\pi)^3} e^{i\textbf{k}\cdot (\textbf{x}-\textbf{y})}  \bigg( \delta_{jk} - \dfrac{k_j k_k}{k^l k_l} \bigg)\, . \\
    \end{split}
\end{equation}
These correlators do not depend on our choice of hydrodynamic frame since $\tau$ does not appear anywhere. This reassures us that they constitute a robust prediction of the model. The Fourier integral of the $\mathcal{E}\mathcal{E}$ correlator may be evaluated analytically. However, it is more illuminating to take the divergence of the electric fields inside the correlator and use the Maxwell equation $\partial_j \mathcal{E}^j=J^0$ to find the density-density correlator. The result is
\begin{equation}\label{gumpfone}
    \langle J^0(\textbf{x})  J^0(\textbf{y}) \rangle= \delta^3(\textbf{x}{-}\textbf{y}) -\dfrac{\Sigma \, e^{-\sqrt{\frac{\Sigma}{D}} |\textbf{x}-\textbf{y}|}}{4\pi D |\textbf{x}-\textbf{y}|} \, .
\end{equation}
The Dirac delta was already present in the theory of diffusion without the electromagnetic coupling; see section \ref{causuzzuz}. The negative Yukawa-type potential is an electromagnetic correction, which incorporates the Debye screening effect (with Debye length $\sqrt{D/\Sigma}$). In fact, the right-hand side of \eqref{gumpfone} coincides with the equilibrium charge-density at $\textbf{x}$ that is generated if we insert a static unit point charge at $\textbf{y}$ (see Appendix \ref{debby} for the proof). 

Let us now focus on the $\mathcal{B}\mathcal{B}$ correlator. If we contract it with a constant vector $m^k$, and we evaluate the Fourier integral explicitly (setting $\textbf{y}=0$ for convenience), we  obtain 
\begin{equation}
\langle\mathcal{B}_j(\textbf{x})\mathcal{B}_k(\textbf{0})m^k \rangle  = \dfrac{ D}{4\pi \Sigma} \bigg[ \dfrac{3x_j (\textbf{x}\cdot \textbf{m})  -|\textbf{x}|^2 m_j}{|\textbf{x}|^5}+ \dfrac{8\pi}{3} m_j \delta^3(\textbf{x}) \bigg] \, .
\end{equation}
The right-hand side is $D/\Sigma$ times the magnetic field induced by an ideal dipole with magnetic moment $m^j$ placed in the origin, see equation (5.64) of \cite{jackson_classical_1999}. Physically, this tells us that if there is a magnetic fluctuation at a point, this is probably generated by a current loop in a neighborhood of such a point, and we can therefore expect to detect a dipole field in the surroundings.

\section{Bulk viscosity in the pressure frame}\label{bulkuzzo}

In this section, we regularize the information current of a simple causal and stable first-order model for the bulk viscosity of a relativistic fluid at zero chemical potential.

\subsection{Overview of the model}

Consider a fluid whose hydrodynamic fields are the temperature $T$ and the flow velocity $u^\mu$. Postulate the following BDNK-type \cite{Bemfica2017TheFirst,GavassinoSymmetric2022,Noronha:2021syv,Bemfica:2022dnk} constitutive relations for the stress-energy tensor and the entropy current:
\begin{equation}
\begin{split}
T^{\mu \nu}={}& (\varepsilon+P+\mathcal{A})u^\mu u^\nu +Pg^{\mu \nu} \, , \\
S^\mu ={}& (s+\mathcal{A}/T)u^\mu \, , \\
\end{split}
\end{equation}
where $\varepsilon(T)$, $s(T)$, and $P(T)$ are the equilibrium energy density, entropy density, and pressure, respectively. They are related by standard thermodynamic identities: $d\varepsilon =Tds=c_v dT$, where $c_v(T)$ is the specific heat, $dP=sdT$, and $Ts= \varepsilon+P$. The scalar field $\mathcal{A}$ is a first-order bulk viscous correction, which is given by
\begin{equation}\label{zimared}
\mathcal{A}=g u^\mu \partial_\mu T + h \partial_\mu u^\mu \, ,
\end{equation}
where $g(T)$ and $h(T)$ are two transport coefficients. If we linearize about homogeneous equilibrium, the linear perturbation fields $\delta T$ and $\delta u^j$ obey the following equations of motion (in the equilibrium rest frame):
\begin{equation}\label{zimablue}
\begin{split}
& c_v \partial_t \delta T +g \partial^2_t \delta T +(\varepsilon{+}P)\partial_j \delta u^j + h\partial_t \partial_j \delta u^j =0 \, , \\
& (\varepsilon{+}P)\partial_t \delta u_j + s \partial_j \delta T =0 \, , \\
\end{split}
\end{equation}
which follow from the linearized conservation law $\partial_\mu \delta T^{\mu \nu}=0$. The above equations can be combined into a single field equation for the temperature perturbation, namely
\begin{equation}\label{gzi}
c_v \partial^2_t \delta T -s \partial_j \partial^j \delta T + g \partial^3_t \delta T -\dfrac{h}{T} \partial_t \partial_j \partial^j \delta T=0 \, .
\end{equation}
Assuming (in accordance with thermodynamics) that $c_v$, $s$, and $T$ are positive, we can derive the conditions for covariant stability (and therefore causality) of the model from equation \eqref{gzi}. Introducing the speed of sound squared $c_s^2=s/c_v$, we have the following inequalities:
\begin{equation}\label{mustuzzo}
0< c_s^2 <1 \, , \spc g > \dfrac{h}{T} > gc_s^2 >0 \, .
\end{equation}

\subsection{Information current and entropy production rate}

Constructing the regularized information current and entropy production rate for the model \eqref{zimablue} requires a bit of trial and error. Here, we only provide the result:
\begin{equation}\label{INFOONA2}
\begin{split}
TE^0 ={}& \dfrac{1}{2} \bigg[ \dfrac{c_v}{T} (\delta T)^2 +(\varepsilon{+}P)\delta u^j \delta u_j +2 \dfrac{\delta T}{T} (g \partial_t \delta T +h \partial_j \delta u^j) + \dfrac{\lambda}{T} (g \partial_t \delta T +h \partial_j \delta u^j)^2 \bigg] \, ,\\
TE^j ={}& s \, \delta T \delta u^j \, ,\\
T\sigma ={}& \dfrac{\lambda s}{h} (g\partial_t \delta T +h \partial_j \delta u^j)^2 \, , \\
\end{split}
\end{equation}
with
\begin{equation}\label{lambdone}
\lambda = \dfrac{1}{\bigg(1- \dfrac{gT}{h} c_s^2 \bigg) c_v} \, .
\end{equation}
To verify that this works, it is enough to write the equation $T\partial_\mu E^\mu +T\sigma=0$ explicitly. This gives
\begin{equation}
\dfrac{1}{T} (\delta T {+}\lambda g \partial_t \delta T {+}\lambda h \partial_j \delta u^j)\big[ c_v \partial_t \delta T+g \partial^2_t \delta T+(\varepsilon{+}P)\partial_j \delta u^j+h \partial_t \partial_j \delta u^j\big]+\delta u^j \big[(\varepsilon{+}P)\partial_t \delta u_j+s\partial_j \delta T \big]=0 \, ,
\end{equation}
which is automatically satisfied along solutions of the equations of motion \eqref{zimablue}.

\subsection{Thermodynamic stability}

Let us now analyze the conditions for thermodynamic stability. Assuming that $T>0$, we can require $TE^0$ and $T\sigma$ to be positive, and this implies (by direct inspection) that $c_v$, $s$, $\lambda$, and $h$ are all positive (us the identity $\varepsilon+P=Ts$), as expected. The fact that $\lambda>0$ implies $h/T > gc_s^2$, in agreement with \eqref{mustuzzo}. Now we only need to find the conditions under which $E^\mu$ is future directed non-spacelike. In particular, introducing the notation $\delta \mathcal{A}=g \partial_t \delta T+h \partial_j \delta u^j$, in accordance with \eqref{zimared}, we have to impose that
\begin{equation}
T(E^0-E^1)=\dfrac{1}{2T} \big( \delta T , \delta u_1, \delta \mathcal{A} \big) \begin{bmatrix}
c_v & -Ts & 1 \\
-Ts & T^2 s & 0 \\
1 & 0 & \lambda \\
\end{bmatrix}
\begin{pmatrix}
\delta T \\
\delta u_1 \\
\delta \mathcal{A} \\
\end{pmatrix}
 + \dfrac{1}{2}(\varepsilon {+}P) \big[(\delta u_2)^2 + (\delta u_3)^2 \big] \geq 0 \, .
\end{equation}
The determinants of the $2\times 2$ blocks of the matrix above lead to the inequalities $\lambda>1/c_v$ and $c_v>s$. The latter can be rewritten in the form $c_s^2<1$. Finally, the positivity of the determinant of the whole $3 \times 3$ matrix produces the more stringent inequality
\begin{equation}\label{lambdino}
\lambda > \dfrac{1}{(1-c_s^2)c_v} \, .
\end{equation}
Combining \eqref{lambdone} and \eqref{lambdino}, we finally obtain the inequality $g>h/T$. Thus, we have recovered all the inequalities of \eqref{mustuzzo}, meaning that hydrodynamic stability and thermodynamic stability are equivalent in this model.

\subsection{Equal-time correlators}

Given the information current reported in equation \eqref{INFOONA2}, we can recast the integral $E=\int E^0 d^3x$ in the form \eqref{mazurca}, where the operator $\mathcal{K}$, acting on the ordered triplet of functions $\{\delta T, \partial_t \delta T, \delta u_j \}$, is provided below:
\begin{equation}
    T^2\mathcal{K}(\partial_j) = 
    \begin{bmatrix}
c_v & g & h \nabla^T \\
g  & \lambda g^2 & \lambda gh \nabla^T \\
-h\nabla & -\lambda gh\nabla & T(\varepsilon{+}P)\mathbb{I}{-}\lambda h^2 \nabla \nabla^T
    \end{bmatrix} \, .
\end{equation}
Using equation \eqref{feynman}, we can compute all the field-field correlators:
\begin{equation}\label{babyfeynman}
\begin{bmatrix}
  \langle \delta T(\textbf{x}) \, \delta T(\textbf{y}) \rangle   &  \langle \delta T(\textbf{x}) \, \partial_t \delta T(\textbf{y}) \rangle & \langle \delta T(\textbf{x}) \, \delta u^T(\textbf{y}) \rangle\\
  \langle \partial_t \delta T(\textbf{x}) \delta T(\textbf{y}) \rangle   &  \langle \partial_t \delta T(\textbf{x}) \partial_t \delta T(\textbf{y}) \rangle & \langle \partial_t \delta T(\textbf{x}) \delta u^T(\textbf{y}) \rangle\\
\langle \delta u(\textbf{x}) \, \delta T(\textbf{y}) \rangle   &  \langle \delta u(\textbf{x}) \, \partial_t \delta T(\textbf{y}) \rangle & \langle \delta u(\textbf{x}) \, \delta u^T(\textbf{y}) \rangle\\
\end{bmatrix}
     = T \! \!\int \dfrac{d^3 k}{(2\pi)^3} e^{i\textbf{k}\cdot (\textbf{x}-\textbf{y})} 
\begin{bmatrix}
\lambda b & {-}\dfrac{b}{g} & 0 \\
{-}\dfrac{b}{g} & \dfrac{bc_v}{g^2}{+}\dfrac{h^2 k^j k_j}{g^2(\varepsilon{+}P)} & {-}\dfrac{ih\textbf{k}^T}{g(\varepsilon{+}P)} \\
0 & \dfrac{ih\textbf{k}}{g(\varepsilon{+}P)} & \dfrac{\mathbb{I}}{\varepsilon{+}P}
     \end{bmatrix},
\end{equation}
where we have introduced the postive parameter $b=T/(c_v \lambda{-}1)$. Clearly, most of these correlators depend on the hydrodynamic frame, being sensitive to both $g$ and $h$. This reflects the fact that the very definitions of $\delta T$ and $\delta u^j$ are related to a specific choice of frame \cite{Kovtun2019}. On the other hand, for the BDNK approach to be reliable, the fluctuations of the energy density $\delta T^{00}=c_v \delta T +g\partial_t \delta T +h\partial_j \delta u^j$ and of momentum density $\delta T^{0j}=(\varepsilon+P)\delta u^j$ should not depend on this choice (at least in the infrared limit). Indeed, this is precisely what happens:
\begin{equation}
\begin{split}
    \langle \delta T^{00}(\textbf{x})\delta T^{00}(\textbf{y})\rangle ={}& c_v T^2 \delta^3(\textbf{x}-\textbf{y}) \, , \\ \langle \delta T^{00}(\textbf{x})\delta T^{0j}(\textbf{y})\rangle ={}& 0 \, ,\\   \langle \delta T^{0j}(\textbf{x})\delta T^{0k}(\textbf{y})\rangle ={}& T(\varepsilon{+}P) \delta^{jk}\delta^3(\textbf{x}-\textbf{y}) \, . \\
    \end{split}
\end{equation}
These correlators agree with the standard equal-time correlators of fluctuating hydrodynamics \cite{landau9}.

\section{Causal shear viscosity}\label{shearuzzuzuzuz}

We consider now a simple first-order hydrodynamic model that also includes shear viscosity.

\subsection{Overview of the model}

Let us consider a viscous fluid at zero chemical potential in the Landau frame. The linearized equations of motion for the temperature and flow velocity fluctuations read
\begin{equation}\label{nobuono}
\begin{split}
& c_v \partial_t \delta T +(\varepsilon{+}P)\partial_j \delta u^j =0 \, ,\\
& (\varepsilon{+}P)\partial_t \delta u_j+s\partial_j \delta T -2 \eta \partial^k \partial_{(k}\delta u_{j)} =0\, , \\
\end{split}
\end{equation}
where $\eta$ is the shear viscosity coefficient. To simplify the equations, we have also added a bulk viscous term with viscosity coefficient $\zeta=2\eta/3$. It is well known that the system \eqref{nobuono} defines an acausal and unstable (when boosted) model. Hence, let us perform a change of hydrodynamic frame of the form $\delta u^j \rightarrow \delta u^j +\tau \partial_t \delta u^j$, for some constant $\tau$. The resulting equations, truncated to second order in derivatives, read
\begin{equation}\label{sistemone}
\begin{split}
& c_v \partial_t \delta T +(\varepsilon{+}P)(\partial_j \delta u^j {+}\tau \partial_t \partial_j \delta u^j )=0 \, ,\\
& (\varepsilon{+}P)(\partial_t \delta u_j{+}\tau \partial_t^2 \delta u_j)+s\partial_j \delta T -2 \eta \partial^k \partial_{(k}\delta u_{j)} =0\, . \\
\end{split}
\end{equation}
This is a simple example of BDNK theory for viscosity. Below, we construct a regularized information current for this model.

\subsection{Duality with Israel-Stewart 
 Theory}

First of all, let us find the conditions under which the system \eqref{sistemone} is causal and stable, assuming that $s$, $c_v$, and $T$ are positive. To this end, we can employ a surprising trick. Let us define the fields
\begin{equation}\label{shangchi}
\begin{split}
\delta v_j ={}& \delta u_j +\tau \partial_t \delta u_j \, , \\
\delta \Pi_{kj}={}& -2\eta \partial_{(k}\delta u_{j)} \, .\\
\end{split}
\end{equation}
Then, by changing variables in system \eqref{sistemone}, we obtain an exact Israel-Stewart model (in agreement with \cite{DoreGavassino2022}):
\begin{equation}\label{sistemoneIS}
\begin{split}
& c_v \partial_t \delta T +(\varepsilon{+}P)\partial_j \delta v^j =0 \, ,\\
& (\varepsilon{+}P)\partial_t \delta v_j+s\partial_j \delta T + \partial^k \delta \Pi_{kj} =0\, , \\
& \tau \partial_t \delta \Pi_{kj}+\delta \Pi_{kj}=-2\eta \partial_{(k} \delta v_{j)} \, . \\
\end{split}
\end{equation}
This establishes a mathematical equivalence between the BDNK model \eqref{sistemone} and the Israel-Stewart model \eqref{sistemoneIS} since all solutions of the former can be converted into (exact) solutions of the latter through the change of variables \eqref{shangchi}\footnote{Note that the reversal is not true: there are solutions of \eqref{sistemoneIS} which do not have a correspondent in \eqref{sistemone}. This is because \eqref{sistemoneIS} possesses 10 mathematical degrees of freedom (namely $\delta T$, $\delta u_j$, $\delta \Pi_{kj}$), while \eqref{sistemone} possesses only 7 (namely $\delta T$, $\delta u_j$, $\partial_t\delta u_j$).}. But the conditions for causality and stability of \eqref{sistemoneIS} are well known (recall that $c_s^2=s/c_v$):
\begin{equation}\label{condazze}
\tau,\eta >0 \, , \spc 0 <c_s^2<1 \, , \spc c_s^2+\dfrac{2\eta}{(\varepsilon {+} P)\tau} <1 \, ,
\end{equation}
and it is straightforward to show that if \eqref{sistemoneIS} is causal and stable, then also \eqref{sistemone} must be causal and stable.

\subsection{Information current and entropy production rate}

We can use the duality with Israel-Stewart theory as a guiding principle for constructing the information current and entropy production rate of system \eqref{sistemone}. In particular, we can first write out the information current and entropy production rate of \eqref{sistemoneIS}, which are well known \cite{GavassinoUniversality2023}, and then use the change of variables \eqref{shangchi} to express the result in terms of the BDNK fields. The result is
\begin{equation}\label{INFOONA3}
\begin{split}
TE^0 ={}& \dfrac{1}{2} \bigg[ \dfrac{c_v}{T} (\delta T)^2 +(\varepsilon{+}P)(\delta u^j{+}\tau \partial_t \delta u^j)(\delta u_j{+}\tau \partial_t  \delta u_j) +2\eta \tau \partial_{(j}\delta u_{k)} \partial^{(j}\delta u^{k)} \bigg] \, ,\\
TE^j ={}& s \, \delta T (\delta u^j{+}\tau \partial_t \delta u^j)-2\eta (\delta u_k {+}\tau \partial_t \delta u_k)\partial^{(k}\delta u^{j)} \, ,\\
T\sigma ={}& 2\eta \partial_{(j}\delta u_{k)} \partial^{(j}\delta u^{k)} \, . \\
\end{split}
\end{equation}
This is indeed the ``correct'' information current, since the equation $T\partial_\mu E^\mu +T\sigma=0$ explicitly reads
\begin{equation}
\dfrac{\delta T}{T} \big[c_v \partial_t \delta T +(\varepsilon{+}P)(\partial_j \delta u^j {+}\tau \partial_t \partial_j \delta u^j ) \big]+(\delta u^j{+}\tau \partial_t \delta u^j)\big[ (\varepsilon{+}P)(\partial_t \delta u_j{+}\tau \partial_t^2 \delta u_j)+s\partial_j \delta T -2 \eta \partial^k \partial_{(k}\delta u_{j)} \big]=0 \, ,
\end{equation}
which is automatically satisfied along exact solutions of the equations of motion \eqref{sistemone}. 

Finally, we can import well-known results of Israel-Stewart theory, and conclude that the conditions \eqref{condazze}, under which \eqref{sistemoneIS} is causal and stable, are also the conditions that make $E^\mu$ timelike future-directed, and $\sigma$ non-negative. 

\subsection{Equal-time correlators}

Since, in the formula for $E^0$, the temperature decouples from the velocity, all temperature-velocity correlators vanish, while the temperature-temperature correlator is just $\langle \delta T(\textbf{x})\delta T(\textbf{y})\rangle= T^2\delta^3(\textbf{x}-\textbf{y})/c_v$, see \cite{landau9}. Let us focus on the velocity correlators. For the ordered list of variables $\{\delta u_j,\partial_t \delta u_j \}$, the integral of $E^0$ takes the form \eqref{mazurca}, with
\begin{equation}
    T\mathcal{K}(\partial_j) = 
    \begin{bmatrix}
        (\varepsilon{+}P{-}\eta\tau \partial_j \partial^j)\mathbb{I}{-}\eta \tau \nabla \nabla^T & & (\varepsilon{+}P) \tau \mathbb{I} \\
        (\varepsilon{+}P) \tau \mathbb{I} & & (\varepsilon{+}P) \tau^2 \mathbb{I} \\
    \end{bmatrix} \, .
\end{equation}
Hence, from equation \eqref{feynman}, we obtain
\begin{equation}
\begin{bmatrix}
     \langle \delta u(\textbf{x}) \, \delta u^T(\textbf{y}) \rangle   &  \langle \delta u(\textbf{x}) \, \partial_t \delta u^T(\textbf{y}) \rangle\\
   \langle \partial_t \delta u(\textbf{x}) \, \delta u^T(\textbf{y}) \rangle  & \langle \partial_t \delta u(\textbf{x}) \, \partial_t \delta u^T(\textbf{y}) \rangle \\
\end{bmatrix}
     = T \! \! \int \dfrac{d^3 k}{(2\pi)^3} \, \dfrac{ e^{i\textbf{k}\cdot (\textbf{x}-\textbf{y})}}{\eta \tau \textbf{k}^2} \, 
     \begin{bmatrix}
         \mathbb{I}{-} \dfrac{\textbf{k}\textbf{k}^T}{2\textbf{k}^2} & {-}\dfrac{1}{\tau} \bigg( \mathbb{I}{-} \dfrac{\textbf{k}\textbf{k}^T}{2\textbf{k}^2}\bigg) \\
         {-}\dfrac{1}{\tau} \bigg( \mathbb{I}{-} \dfrac{\textbf{k}\textbf{k}^T}{2\textbf{k}^2}\bigg) &  \dfrac{1}{\tau^2} \bigg( \dfrac{\varepsilon{+}P{+}\eta \tau \textbf{k}^2}{\varepsilon{+}P}\mathbb{I}{-} \dfrac{\textbf{k}\textbf{k}^T}{2\textbf{k}^2}\bigg)\\
     \end{bmatrix} \, . 
\end{equation}
It is evident that these correlators are not invariant under a change of hydrodynamic frame since the UV-cutoff parameter $\tau$ appears explicitly. This is not a surprise because $\delta u^j$ is not a physically relevant quantity, being ``hydrodynamic frame-dependent''. To check whether the \textit{physical predictions} of the model are sensitive to the hydrodynamic frame, we need to compute the correlator of the momentum density $\delta T^{0j}=(\varepsilon{+}P)(\delta u^j {+}\tau \partial_t \delta u^j)$, which is a ``hydrodynamic frame-invariant'' observable. Interestingly, we find that
\begin{equation}
    \langle \delta T^{0j}(\textbf{x})\delta T^{0k}(\textbf{y})\rangle = T(\varepsilon{+}P)\delta^{jk}\delta^3(\textbf{x}-\textbf{y}) \, ,
\end{equation}
which does not depend on $\tau$ and is the natural relativistic generalization of equation (88.5) of \cite{landau9}.

\section{Goldstone modes} \label{tinytim}

Up to this point, we have constructed information currents for first-order dissipative models. However, there are also first-order theories that describe non-dissipative systems (i.e. systems with $\sigma=0$).  This is the case, for example, when the field $\varphi$ is a Goldstone mode arising from spontaneous symmetry breaking. In this setting, the dynamics must be invariant under global shifts $\varphi \rightarrow \varphi+a$ (with $a=\text{const}$), meaning that $E^\mu$ can depend only on $\partial \varphi$, and not on $\varphi$ itself. We shall discuss below a couple of examples.

\subsection{U(1) symmetry breaking}\label{SySy}

Let $\varphi$ be the perturbation to the phase field arising from a $U(1)$ spontaneous symmetry breaking. Since Goldstone modes are massless, in the linear regime, the equation of motion for $\varphi$ is a wave-type equation:
\begin{equation}\label{moMMU}
\partial^2_t \varphi -c_s^2 \partial_j \partial^j \varphi=0 \, ,
\end{equation}
where $c_s$ is the speed of the associated phonon-type excitation. This model can be viewed as a first-order theory because the equation of motion can be rewritten in the form of a conservation law, $\partial_\mu J^\mu=0$, with $J^\mu =(-\partial_t \varphi,c_s^2 \partial^j \varphi)$, in accordance with zero-temperature superfluid dynamics \cite{cool1995,Termo,GavassinoStabilityCarter2022}. Causality and stability demand $0<c_s^2<1$. The information current of this system (possibly rescaled by some overall constant) is
\begin{equation}\label{uuuno}
\begin{split}
E^0 ={}& \dfrac{1}{2}\big[(\partial_t \varphi)^2+c_s^2 \partial_j \varphi \partial^j \varphi \big] \, , \\
E^j ={}& -c_s^2 \partial_t \varphi \partial^j \varphi \, . \\
\end{split}
\end{equation}
In fact, the equation $\partial_\mu E^\mu =0$ explicitly reads
\begin{equation}
\partial_t \varphi (\partial^2_t \varphi -c_s^2 \partial_j \partial^j \varphi)=0 \, ,
\end{equation}
which is automatically satisfied along solutions of the equations of motion \eqref{moMMU}. Not surprisingly, $E^\mu$ is timelike future directed if and only if $0<c_s^2<1$.

Now we can compute equal time correlators. From equation \eqref{uuuno}, we obtain
\begin{equation}
    \mathcal{K}(\partial_j)= 
    \begin{bmatrix}
      -c_s^2 \partial_j \partial^j  &  0 \\
       0 &  1 \\
    \end{bmatrix} \, ,
\end{equation}
so that we have
\begin{equation}\label{er9ofmel}
\begin{bmatrix}
      \langle \varphi(\textbf{x}) \, \varphi(\textbf{y}) \rangle   &  \langle \varphi(\textbf{x}) \, \partial_t \varphi(\textbf{y}) \rangle\\
   \langle \partial_t \varphi(\textbf{x}) \, \varphi(\textbf{y}) \rangle  & \langle \partial_t  \varphi(\textbf{x}) \, \partial_t \varphi(\textbf{y}) \rangle \\
\end{bmatrix}
     = \int \dfrac{d^3 k}{(2\pi)^3} \, e^{i\textbf{k}\cdot (\textbf{x}-\textbf{y})} \, \begin{bmatrix}
         (c_s^2 k^j k_j)^{{-}1} & & 0 \\
         0 & & 1 \\
     \end{bmatrix} \, .
\end{equation}
Let us note that, by the Josephson-Anderson relation, we have $\partial_t \varphi \propto \delta \mu$, where $\mu$ is chemical potential. Hence, by integrating the Fourier integral above, we find that $\langle \delta \mu(\textbf{x})\delta \mu(\textbf{y})\rangle\propto \delta^3(\textbf{x}-\textbf{y})$, which again is consistent with the analysis of section \S 88 of \cite{landau9}. Interestingly enough, the correlator of the phase field with itself is long-range, since the integral of \eqref{er9ofmel} gives (in agreement with \cite{landau9}, \S 87, Problem 2)
\begin{equation}\label{U1U}
     \langle \varphi(\textbf{x}) \, \varphi(\textbf{y}) \rangle  \propto \dfrac{1}{ |\textbf{x}-\textbf{y}|} \, .
\end{equation}
This is due to the fact that long-wavelength fluctuations are very likely to occur in this system. In fact, the free energy cost of a perturbation to $\varphi$ scales like $\partial^2$, meaning that $e^{-E}\rightarrow 1$ in the IR limit, inducing correlations over large distances.

\subsection{Elastic media}

Let $\xi_j$ be the linear displacement vector field of an elastic isotropic material. This quantifies the departure of the material elements from the unstrained (i.e. equilibrium) configuration. Clearly, a global shift $\xi_j \rightarrow \xi_j +a_j$ (with $a_j=\text{const}$) cannot result in a change of entropy, since we are translating the whole system rigidly. Hence, one can treat $\xi_j$ as a Goldstone-type effective field. The equation of motion, as predicted by the theory of elasticity in the linear regime \cite{landau7}, reads
\begin{equation}\label{EUOM}
\partial^2_t \xi_j - \mu \partial^k \partial_k \xi_j- (\mu {+}\lambda)\partial_j \partial^k \xi_k =0 \, ,
\end{equation}
where $\mu$ and $\lambda$ are, respectively, the shear modulus and Lam\'e{}'s first parameter, both rescaled by the enthalpy density. The conditions for causality and stability of this model are $0 < \mu <1$ (causality and stability of shear waves) and $0<2\mu {+}\lambda<1$ (causality and stability of sound waves). The information current, rescaled by some background constant, is 
\begin{equation}
\begin{split}
E^0 ={}& \dfrac{1}{2} \big[\partial_t \xi_j \partial_t \xi^j +2\mu \partial_{(j}\xi_{k)}\partial^{(j}\xi^{k)}+\lambda (\partial_j \xi^j)^2 \big] \, ,\\
E^j ={}& -2\mu \partial_t \xi_k  \partial^{(k} \xi^{j)}-\lambda \partial_k \xi^k \partial_t \xi^j \, . \\
\end{split}
\end{equation}
In fact, the equation $\partial_\mu E^\mu=0$ explicitly reads
\begin{equation}
\partial_t \xi^j \big[ \partial^2_t \xi_j -\mu \partial^k \partial_k \xi_j -(\mu{+}\lambda)\partial_j \partial^k \xi_k \big]=0 \, ,
\end{equation}
which is automatically satisfied along all solutions of the equation of motion \eqref{EUOM}. The necessary and sufficient conditions for $E^\mu$ to be timelike future directed are reported below:
\begin{equation}
  0 < \mu <1 \, , \spc  0<2\mu {+}\lambda<1 \, , \spc \lambda+\dfrac{2}{3} \mu >0 \, . 
\end{equation}
The first two bounds are just the same causality and stability conditions that we derived from the equation of motion \eqref{EUOM}. The third condition is the thermodynamic requirement that the bulk modulus should be non-negative \cite{landau7}. 

The computation of the equal-time correlators is, again, straightforward. The velocity-velocity correlator is $\langle \partial_t \xi^j(\textbf{x}) \partial_t \xi^k(\textbf{y}) \rangle=\delta^{jk}\delta^3(\textbf{x}-\textbf{y})$, the velocity-displacement correlator vanishes, and the displacement-displacement correlator reads
\begin{equation}
   \langle \xi^j(\textbf{x})  \xi^k(\textbf{y}) \rangle = \int \dfrac{d^3 k}{(2\pi)^3} \, \dfrac{e^{i\textbf{k}\cdot (\textbf{x}-\textbf{y})}}{\textbf{k}^2} \bigg[ \dfrac{\delta^{jk}}{\mu}  +\bigg( \dfrac{1}{2\mu +\lambda} -\dfrac{1}{\mu} \bigg) \dfrac{k^j k^k}{k^l k_l}  \bigg] \, .
\end{equation}
This Fourier integral simplifies considerably if we contract it with $\delta_{jk}$. The result is
\begin{equation}
     \langle \xi^j(\textbf{x})  \xi_j(\textbf{y}) \rangle = \bigg( \dfrac{2}{\mu}{+}\dfrac{1}{2\mu {+}\lambda} \bigg) \dfrac{1}{4\pi |\textbf{x}-\textbf{y}|} \, .
\end{equation}
One can see that the field-field correlator scales with the inverse of the distance. This is a common property of all Goldstone modes (in $3{+}1$ dimensions), as testified also by Eq.\ \eqref{U1U}. This universal $1/r$ behavior is a direct consequence of Goldstone's theorem \cite{huang_book,Zee2003}.

\section{Chiral hydrodynamics}\label{CHIRUZ}

Another example of a non-dissipative first-order hydrodynamic theory is ideal chiral hydrodynamics 
\cite{Erdmenger:2008rm, Banerjee:2008th, Son:2009tf,Sadofyev:2010pr,Neiman:2010zi,Landsteiner:2012kd,Chen:2015gta,Speranza2021}. Let us construct its regularized information current using our new method.

\subsection{Overview of the model}\label{overv}

The fields of ideal chiral hydrodynamics are the same as those of ideal hydrodynamics at finite chemical potential: $T$ (temperature), $\mu$ (chemical potential), and $u^\mu$ (flow velocity). However, in contrast to ideal hydrodynamics, the chemical potential $\mu$ is associated with a pseudoscalar charge, i.e. the charge density $J^0$ is odd under spatial inversion and the current density $J^i$ is a pseudovector. This difference allows for additional dissipationless terms in the constitutive relations for the stress-energy tensor, the particle current, and the entropy current:
\begin{equation}\label{CHIR}
    \begin{split}
T^{\mu \nu}={}& (\varepsilon{+}P)u^\mu u^\nu +Pg^{\mu \nu}+\xi_T (u^\mu \omega^\nu{+}u^\nu \omega^\mu) \, , \\
J^{\mu}={}& nu^\mu +\xi_J \omega^\mu \, , \\
S^{\mu}={}& su^\mu +\xi_S \omega^\mu \, , \\
    \end{split}
\end{equation}
where $\omega^\mu = \frac{1}{2}\varepsilon^{\mu \nu \rho \sigma}u_\nu \partial_\rho u_\sigma$ is the kinematic vorticity vector. As usual, $\varepsilon$, $P$, $n$, and $s$ are the energy density, pressure, particle density, and entropy density, respectively. Standard thermodynamic relations hold: $d\varepsilon=Tds+\mu dn$, $dP=sdT+nd\mu$, and $\varepsilon{+}P=Ts{+}\mu n$. The parameters $\xi_T$, $\xi_J$, and $\xi_S$ are transport coefficients encoding the presence of the chiral anomaly that depend on the choice of hydrodynamic frame (namely, on the definition of $u^\mu$) \cite{Son:2009tf,Rajagopal:2015roa, Stephanov:2015roa, Chen:2015gta, Yang:2018lew,Speranza2021}.

Considering linearized dynamics around a homogeneous (non-rotating) equilibrium, the conservation laws $\partial_\mu T^{\mu \nu}=0$ and $\partial_\mu J^\mu=0$ can be written as follows:
\begin{equation}\label{beatles}
    \begin{split}
& \partial_t \delta s +s \partial_j \delta u^j =0 \, , \\
& \partial_t \delta n +n \partial_j \delta u^j =0 \, ,\\
& (\varepsilon{+}P)\partial_t \delta u_j +\partial_j \delta P - \dfrac{1}{2}\xi_T \partial_t (\nabla {\times} \delta u)_j =0 \, .\\
    \end{split}
\end{equation}
It is straightforward to verify that if $\xi_T \neq 0$ the system is unstable. In fact, it admits solutions of the form
\begin{equation}\label{pathuz}
\begin{split}
   & \delta T = \delta \mu =0 \, , \\
    & \delta u_j =f(t) \big( \cos(az), \,  -\sin(az), \, 0 \big) \, , \spc \text{with }\, a= \dfrac{2(\varepsilon+P)}{\xi_T} \, , \\
\end{split}
\end{equation}
for arbitrary $f(t)$. Since we can set $f(t)=e^{t}$, we see that there are growing Fourier modes, manifesting an instability. This implies that there cannot be an information current with the properties listed in \cite{GavassinoCausality2021}, as these would imply stability. We note that the linearized system is actually ``infinitely unstable'', since the growth of $f(t)$ can be arbitrarily fast, and there is no Lyapunov exponent controlling the explosion. This is a signature of Hadamard's ill-posedness, as infinitely small initial conditions at $t=0$ can give rise to infinite large solutions at $t=0^+$ \cite{Rauch_book}. Indeed, the ill-posedness could directly be argued from the non-uniqueness of $f(t)$, which tells us that the initial value problem is non-deterministic. The same conclusion about the ill-posedness was also found in \cite{Speranza2021}. Physically, the solution \eqref{pathuz} represents a circularly polarized shear wave, which can grow in size while conserving the momentum density $\delta T^{0j}=(\varepsilon{+}P)\delta u^j-\frac{1}{2}\xi_T (\nabla \times \delta u)^j$ through a careful cancellation. This is completely analogous to the instability of the Eckart theory, where the fluid uses the first-order correction to the momentum (the heat flux, in the case of Eckart) as ``rocket fuel'' to accelerate at constant momentum density \cite{GavassinoLyapunov_2020}.

For the reasons above, in the following, we set $\xi_T=0$ through the field redefinition $\delta u \rightarrow \delta u +a^{-1} \nabla {\times} \delta u$. In the resulting hydrodynamic frame, the system \eqref{beatles} becomes indistinguishable from an ideal non-chiral fluid\footnote{Note that this is true only for non-rotating backgrounds.}.  Still, the entropy and particle currents \eqref{CHIR} have chiral first-order corrections. Hence, even if the linearized equations of motion do not present any chiral features, the information current will exhibit some.

\subsection{First-order information current}

If one starts from the constitutive relations \eqref{CHIR} with $\xi_T=0$, and computes the information current using conventional techniques \cite{GavassinoGibbs2021,GavassinoGENERIC2022}, they obtain the following result, which contains some explicit first-order gradient corrections:
\begin{equation}\label{grzi}
    \begin{split}
TE^0 ={}& \dfrac{1}{2} \bigg[ \delta T \delta s+\delta \mu \delta n + (\varepsilon {+} P)\delta u^j \delta u_j+2\bar{\xi} \delta u^j (\nabla {\times} \delta u)_j \bigg] \, , \\
TE^j ={}& \delta P \delta u^j + \dfrac{2\bar{\xi}}{\varepsilon{+}P} \delta P (\nabla {\times} \delta u)^j + \bar{\xi} (\delta u {\times} \partial_t \delta u)^j \, , \\
    \end{split}
\end{equation}
with
\begin{equation} \label{xibar}
    \bar{\xi}=\dfrac{1}{2} (T\xi_S{+}\mu \xi_J) \, .
\end{equation}
The full derivation is provided in appendix \ref{twonicks}. The reversibility equation $\partial_\mu E^\mu=0$ takes the explicit form
\begin{equation}
    \delta T (\partial_t \delta s +s \partial_j \delta u^j) + \delta \mu (\partial_t \delta n +n\partial_j \delta u^j)+ \bigg[ \delta u^j + \dfrac{2\bar{\xi}}{\varepsilon{+}P} (\nabla {\times} \delta u)^j \bigg] \bigg[ (\varepsilon{+}P)\partial_t \delta u_j + \partial_j \delta P \bigg]=0 \, ,
\end{equation}
which is automatically obeyed along all solutions of the equations of motion \eqref{beatles} (recall that we have set $\xi_T=0$).

\subsection{UV-regularized information current}

Clearly, \eqref{grzi} does not define a timelike future-directed vector. In fact, there are two problems. First of all, in the formula for $E^0$, there is the product $\delta u^j (\nabla {\times} \delta u)_j$, but the term $|\nabla {\times} \delta u|^2$ is missing. Secondly, in the formula for $E^j$, there is a term containing $\partial_t \delta u$, but there is no term $|\partial_t \delta u|^2$ in $E^0$. Luckily, both problems can be fixed.

Let us note that, since $\xi_T=0$, the third equation of \eqref{beatles} implies that $\partial_t (\nabla {\times} \delta u)=0$. It follows that if we add a second-order term $\propto |\nabla {\times} \delta u|^2$ in $E^0$, the equation $\partial_\mu E^\mu =0$ is unchanged, meaning that the resulting second-order information current is still consistent with the equations of motion. 

To solve the problem with $\partial_t \delta u$ inside the formula of $E^j$, we can again use the third equation of \eqref{beatles} with $\xi_T=0$ to derive the following chain of identities:
\begin{equation}
  \delta u \times \partial_t \delta u = -\delta u \times \dfrac{\nabla \delta P }{\varepsilon{+}P} = \nabla \times \bigg( \dfrac{\delta P \delta u}{\varepsilon{+}P} \bigg) -\dfrac{\delta P}{\varepsilon{+}P} \nabla \times \delta u \, .
\end{equation}
We note that, on the rightmost side, the pure curl is a term with vanishing divergence. Hence, if we remove it, the equation $\partial_\mu E^\mu =0$ will still hold \textit{exactly}. Furthermore, removing such a term does not affect the value of the total first-order information at $t=\text{const}$, namely $E=\int E^0 d^3 x$, because it only enters the formula for the flux $E^j$.\footnote{Indeed, this pure curl term does not affect $E$ in any reference frame. In fact, expressed in a covariant language, this term is $\propto \partial_\mu Z^{[\mu \nu]}$, with $Z^{[\mu \nu]}=\varepsilon^{\mu \nu \rho \sigma} u_\rho \delta u_\sigma \delta P $, and thus becomes a boundary term on any Cauchy surface, by Gauss' theorem \cite{Poisson_notes}.}

Hence, with the two modifications above, we obtain the following UV-regularized information current:
\begin{equation}\label{grziUV}
    \begin{split}
TE^0 ={}& \dfrac{1}{2} \bigg[ \delta T \delta s+\delta \mu \delta n + (\varepsilon {+} P)\delta u^j \delta u_j+2\bar{\xi} \delta u^j (\nabla {\times} \delta u)_j +\lambda (\nabla {\times} \delta u)^j  (\nabla {\times} \delta u)_j  \bigg] \, , \\
TE^j ={}& \delta P \delta u^j + \dfrac{\bar{\xi}}{\varepsilon{+}P} \delta P (\nabla {\times} \delta u)^j  \, . \\
    \end{split}
\end{equation}
As expected, the equation $\partial_\mu E^\mu=0$ is exactly consistent with the equations of motion, since it explicitly reads
\begin{equation}
    \delta T (\partial_t \delta s +s \partial_j \delta u^j) + \delta \mu (\partial_t \delta n +n\partial_j \delta u^j)+ \bigg[ \delta u^j + \dfrac{\bar{\xi}}{\varepsilon{+}P} (\nabla {\times} \delta u)^j \bigg] \bigg[ (\varepsilon{+}P)\partial_t \delta u_j + \partial_j \delta P \bigg]+ \big[ \bar{\xi}\delta u^j +\lambda (\nabla \times \delta u)^j\big]\partial_t (\nabla {\times}\delta u)_j=0 \, .
\end{equation}

\subsection{Thermodynamic stability}

We only need to find the conditions under which $E^0-E^1$ is always non-negative. Introducing the notation
\begin{equation}
    \delta \bar{\omega}^j = \dfrac{\bar{\xi}}{\varepsilon{+}P} (\nabla {\times} \delta u)^j \, ,
\end{equation}
we have the following formula
\begin{equation}
    2T(E^0-E^1)= \dfrac{nT}{c_p} \delta \mathfrak{s}^2 +\big(\delta P, \delta u_1, \delta \bar{\omega}_1  \big)
    \begin{bmatrix}
\dfrac{1}{c_s^2(\varepsilon{+}P)} & -1 & -1 \\
-1 & \varepsilon{+}P & \varepsilon{+}P \\
-1 & \varepsilon{+}P & \dfrac{\lambda (\varepsilon{+}P)^2}{\bar{\xi}^2} \\
    \end{bmatrix}
    \begin{pmatrix}
        \delta P \\
        \delta u_1 \\
        \delta \bar{\omega}_1
    \end{pmatrix}
    + \sum_{j=2,3}
    \big(\delta u_j, \delta \bar{\omega}_j  \big)
    \begin{bmatrix}
 \varepsilon{+}P & \varepsilon{+}P \\
\varepsilon{+}P & \dfrac{\lambda (\varepsilon{+}P)^2}{\bar{\xi}^2} \\
    \end{bmatrix}
    \begin{pmatrix}
        
        \delta u_j \\
        \delta \bar{\omega}_j
    \end{pmatrix},
\end{equation}
where $c_p$ is the specific heat at constant pressure, and $\mathfrak{s}$ is the specific entropy. We can then derive the following list of inequalities (assuming that $T$ and $n$ are positive):
\begin{equation}
    \begin{split}
& c_p >0 \, , \\
& \varepsilon{+}P>0 \, , \\
&  0<c_s^2<1 \, , \\
& \lambda > \dfrac{\bar{\xi}^2}{\varepsilon{+}P} \, . \\
    \end{split}
\end{equation}
The first three are needed to guarantee that the ideal fluid equations are causal and stable and enforce textbook thermodynamic inequalities \cite{landau5,Hishcock1983}. The last inequality is needed to guarantee that the chiral sector is stable against stochastic fluctuations at large gradients (i.e. in the UV regime).

\subsection{Physical difference with ideal hydrodynamics}\label{bellalafisica}

In section \ref{overv}, we showed that the linearized equations of motion of ideal chiral hydrodynamics (with $\xi_T=0$) are indistinguishable from those of an ideal fluid if the background state is non-rotating. On the other hand, the information current has a chiral correction, which is absent in non-chiral fluids. This implies that, even though the dynamics is the same, the probability distribution of fluctuations $e^{-E}$ is different between the chiral and the non-chiral case (i.e., the off-shell behavior is sensitive to the presence of the chiral anomaly, even though the on-shell physics is not). Let us analyze such differences in more detail.

We consider the following stationary solutions of the linearized equations of motion:
\begin{equation}\label{circuz}
\begin{split}
   & \delta T = \delta \mu =0 \, , \\
    & \delta u_j =\delta v \big( \cos(kz), \,  \pm\sin(kz), \, 0 \big) \, . \\
\end{split}
\end{equation}
These constitute circularly polarized shear waves of amplitude $\delta v$. The $\pm$ sign defines the handedness of the polarization. Plugging these solutions into \eqref{grziUV}, we obtain the following formula for the information density:
\begin{equation}\label{wmn}
    TE^0 = \dfrac{1}{2} (\varepsilon{+}P) (\delta v)^2 \bigg[ 1 \mp \dfrac{2\bar{\xi} k}{\varepsilon{+}P}+\dfrac{\lambda k^2}{\varepsilon{+}P} \bigg] \, . 
\end{equation}
This tells us that the free energy of a circularly polarized shear wave has a chiral correction (the second term in the square bracket), whose sign depends on the handedness of the wave. Hence, the stochastic fluctuations of a chiral fluid violate parity because right-handed waves have a different probability of being excited with respect to left-handed waves\footnote{Note that, for this effect to exist, we need to have $\bar{\xi} \neq 0$, which means that the background state needs to violate parity in the first place (by possessing a net axial charge). Therefore, the existence of parity-breaking fluctuations is not unexpected. The real surprise is that, while stochastic fluctuations have chiral corrections, these do not survive the deterministic limit (in the linear regime).}.  Note that Eq. \eqref{wmn} is reliable only up to the first order in $k$ because the term proportional to $k^2$ is a mere UV regulator, which was introduced in \eqref{grziUV} to prevent $E^0$ from becoming negative. 

As a next step, one could expand the theory discussed in this section by including the chiral magnetic effect, namely the contribution to the current along the magnetic field \cite{Kharzeev:2015znc}. However, it is important to note that the state with constant $\mu\neq 0$, which implies $\bar{\xi} \neq 0$, is not a true equilibrium state when the magnetic field is treated dynamically, as it is unstable to the formation of electromagnetic waves \cite{Akamatsu:2013pjd, Manuel:2015zpa, Hirono:2015rla,Avdoshkin:2014gpa, Shovkovy:2021yyw}. The extension of the information current for first-order chiral magnetohydrodynamics will be studied in future work.

\subsection{Equal time correlators}

Here we compute only the velocity-velocity correlator of chiral hydrodynamics. In fact, we note that the probability distribution of $\delta T$ and $\delta \mu$ decouples from that of $\delta u^j$ (see equation \eqref{grziUV}). Therefore, we can focus on the velocity, ignoring the other variables (whose fluctuations are standard \cite{landau5}). With a bit of algebra, one can show that for the ordered triplet of variables $(\delta u_1,\delta u_2,\delta u_3)$, the relevant differential operator is given by
\begin{equation}
    T\mathcal{K}(\partial_j) = (\varepsilon{+}P{-}\lambda \partial_j \partial^j)\mathbb{I} + 2\bar{\xi}
    \begin{bmatrix}
        0 & -\partial_3 & \partial_2 \\
       \partial_3 & 0 & -\partial_1 \\
       -\partial_2 & \partial_1 & 0 \\
    \end{bmatrix}+
    \lambda \nabla \nabla^T \, ,
\end{equation}
where $\mathbb{I}$ is the $3{\times}3$ identity matrix. We note that, since there are no time derivatives of the velocity in the information current \eqref{grziUV}, we did not need to include $\partial_t \delta u_j$ among the degrees of freedom. Indeed, this is consistent with the fact that the equations of motion \eqref{beatles} are of first order in time. The equal-time correlator of the velocity then reads
\begin{equation}
\begin{split}
        &\langle \delta u^j(\textbf{x})\delta u^k (\textbf{y})\rangle = {T} \! \! \int \dfrac{d^3 k}{(2\pi)^3} e^{i\textbf{k}\cdot (\textbf{x}-\textbf{y})} G^{jk}(\textbf{k}) \, , \\
        &\text{with } G(\textbf{k}) =\dfrac{\varepsilon{+}P{+}\lambda \textbf{k}^2}{(\varepsilon{+}P{+}\lambda \textbf{k}^2)^2{-}4\bar{\xi}^2\textbf{k}^2} \bigg(\mathbb{I}{-}\dfrac{\textbf{k}\textbf{k}^T}{\textbf{k}^2} \bigg) + \dfrac{2i\bar{\xi}}{(\varepsilon{+}P{+}\lambda \textbf{k}^2)^2{-}4\bar{\xi}^2\textbf{k}^2} \begin{bmatrix}
        0 & k_3 & -k_2 \\
       -k_3 & 0 & k_1 \\
       k_2 & -k_1 & 0 \\
    \end{bmatrix} + (\varepsilon{+}P)^{-1} \dfrac{\textbf{k}\textbf{k}^T}{\textbf{k}^2} \, . \\
\end{split}
\end{equation}
This expression is quite complicated. However, we can make some interesting observations. First of all, let us note that if we set $\bar{\xi}=\lambda=0$, then we recover the ideal fluid formula $\langle \delta u^j(\textbf{x})\delta u^k (\textbf{y})\rangle=T(\varepsilon{+}P)^{-1}\delta^{jk}\delta^3(\textbf{x}{-}\textbf{y})$, which is the relativistic generalization of the equal time velocity-velocity correlator provided in \S 88 of \cite{landau9}. This corresponds to the zeroth order result of the chiral gradient expansion. If now we turn on first-order corrections and take for clarity $\textbf{k}=(k,0,0)$, we find
\begin{equation}
    G(\textbf{k}) =(\varepsilon{+}P)^{-1} \begin{bmatrix}
      1 & 0 & 0 \\
       0 & 1 & \dfrac{2i\bar{\xi}k}{\varepsilon{+}P} \\
       0 & -\dfrac{2i\bar{\xi}k}{\varepsilon{+}P}  & 1 \\ 
    \end{bmatrix}+\mathcal{O}(k^2) \, .
\end{equation}
Therefore, the deviations from the ideal non-chiral fluid are transversal to the wavevector. This is in agreement with our result in Section \ref{bellalafisica} where it was found that chiral corrections affect the probability distribution of circularly polarized waves in a way that depends on their handedness. This suggests that it may be interesting to look at the equal time correlation between the velocity and its curl, whose formula is reported below:
\begin{equation}\label{jkm}
    \langle \delta u^j(\textbf{x}) \, (\nabla {\times} \delta u)_j(\textbf{y})\rangle  = - {T} \! \! \int \dfrac{d^3 k}{(2\pi)^3} e^{i\textbf{k}\cdot (\textbf{x}-\textbf{y})} \dfrac{4\bar{\xi}\textbf{k}^2}{(\varepsilon{+}P{+}\lambda \textbf{k}^2)^2{-}4\bar{\xi}^2\textbf{k}^2} \, .
\end{equation}
One can see that the probability imbalance between left-handed waves and right-handed waves causes the fluid to have a net ``helicity'', which is roughly proportional to the first-order transport coefficient $\bar{\xi}$. Unfortunately, the final formula depends explicitly on the UV-regulator $\lambda$. Not surprisingly, this happens at large $\textbf{k}$, namely outside the regime of applicability of the theory. Hence, if we Fourier-transform both sides, we can derive a useful ``Kubo-like'' formula:
\begin{equation}
    T\xi_S {+}\mu \xi_J = - \lim_{\textbf{k}\rightarrow 0}\dfrac{(\varepsilon{+}P)^2}{2T \textbf{k}^2} \int \langle \delta u^j(\textbf{0}) \, (\nabla {\times} \delta u)_j(\textbf{y})\rangle \, e^{i\textbf{k}\cdot \textbf{y}} d^3 y \, .
\end{equation}

\section{Conclusions} \label{conk}

The information current has a handful of notable applications within relativistic hydrodynamics. It is the most direct means by which we can rigorously establish stability \cite{GavassinoGibbs2021}, causality \cite{GavassinoCausality2021}, and symmetric hyperbolicity \cite{GavassinoNonHydro2022,GavassinoUniversality2023} of the linearized equations of motion of a relativistic fluid model. It can also be used to determine whether two hydrodynamic theories can be mapped into one another  \cite{GavassinoGENERIC2022} and to group large numbers of theories into universality classes \cite{GavassinoUniversality2023}. Additionally, the information current has been employed to derive Onsager-Casimir relations \cite{GavassinoSymmetric2022} and to study hydrodynamic fluctuations both within the Fox-Uhlenbeck \cite{doi:10.1063/1.1693183,Mullins:2023tjg} and the Martin-Siggia-Rose \cite{Martin:1973zz,Mullins:2023ott} approaches. Given all these exciting new perspectives, it was quite disappointing to discover that the standard information current of first-order viscous hydrodynamics  (as computed from the Gibbs criterion \cite{GavassinoGibbs2021}) is always ill-behaved \cite{GavassinoLyapunov_2020}. 
However, after a closer inspection, one realizes that the existence of pathologies is inherent to the derivative expansion, and one always needs to make some adjustments afterwards. For example, the second-order in gradients BRSSS theory \cite{Baier:2007ix} is unstable without a proper resummation which converts it into an Israel-Stewart-like theory. Furthermore, we note that DNMR \cite{Denicol2012Boltzmann} is also unstable if one does not neglect the Knudsen ``$\mathcal{K}$ terms'' (or reabsorbs them through the IReD procedure \cite{WagnerIReD2022}). In BDNK theory one needs to tune the hydrodynamic frame to recover causality and stability \cite{Bemfica2017TheFirst}. In a sense, these strategies are always introduced \textit{a posteriori} to keep the UV sector under control, leaving the IR limit well behaved.

In this work, we have shown that a similar regularization strategy can also be applied to fix the properties of the information current in first-order theories. Unfortunately, so far we could not find a fully general regularization procedure applicable to all theories, but we considered 7 selected models spanning a large spectrum of physical phenomena, for which we always managed to find a successful UV regularization. In some cases, we found that such regularization is not unique, either because it depends on an additional free parameter (see, e.g., section \ref{bazinga!}), or because there are two structurally different regularization schemes (see section \ref{Electro!})\footnote{Strictly speaking, the latter case can always be reduced to the former. In fact, consider two alternative choices of information currents and entropy production rates, $\{E^\mu_1,\sigma_1\}$ and $\{E^\mu_2,\sigma_2\}$. Then, any convex combination $\{(1{-}q)E^\mu_1+qE^\mu_2, (1{-}q)\sigma_1+q\sigma_2\}$, with $q \in [0,1]$, is an equally acceptable choice of information current and entropy production rate. In fact, the properties (i,ii,iii) outlined in the introduction are conserved under convex combinations. More in general, if there are $n$ linearly independent choices of information currents, then there is an $(n{-}1)$-parameter family of convex combinations, all which are equally good choices.}. In other cases, we could find one and only one regularized information current. For example, in the BDNK model for bulk viscosity discussed in Section \ref{bulkuzzo}, we believe that \eqref{INFOONA2} might be the only possible UV-regularized information current, although we cannot present a proof yet. The problem of determining the information current is so highly constrained because we require equation $\partial_\mu E^\mu +\sigma=0$ to hold \textit{exactly} along all solutions of the equations of motion. One may decide to relax this assumption, and demand that $\partial_\mu E^\mu+\sigma=0$ hold only in the IR regime. However, this would destroy the mathematical connection between the information current and causality \cite{GavassinoCausality2021}, possibly introducing observer-dependent pathologies in the stochastic theory \cite{GavassinoSuperluminal2021}.

We would like to remark that the uniqueness of the information current does not imply that there is no freedom in choosing a cutoff scale for our theory. Rather, it implies that the cutoff scale of stochastic fluctuations must match the cutoff scale of deterministic fluctuations. For example, take the BDNK model for shear viscosity discussed in Section \ref{shearuzzuzuzuz}. There, the equations of motion break down on scales comparable to relaxation time $\tau$, see \eqref{sistemone}.  The same relaxation time appears also in the regularized information current \eqref{INFOONA3}. Therefore, the probability distribution $e^{-E}$ for stochastic fluctuations breaks down on the same scale $\tau$ as the equations of motion. Interestingly, in those (rare) cases where the information current depends on a new free parameter, causality and stability always demand that the new cutoff scale associated with this parameter be always greater or equal to the cutoff scale of the equations of motion. Hence, the stochastic theory cannot be applicable in regimes where the deterministic theory is not applicable.

We employed our newly discovered information currents to compute equal-time two-point correlators, and the results are very encouraging. In fact, in our 7 examples, we recovered all the expected physics, including Debye screening, the Wiedemann–Franz law, dipolar magnetic fluctuations, long-range correlations of Goldstone degrees of freedom, and spontaneous helicity generation in parity-violating fluids. In the case of BDNK theories, we noticed an interesting pattern: the equal-time correlators of quantities that depend on the choice of hydrodynamic frame (e.g., temperature, chemical potential, and flow velocity) are very different from their non-relativistic counterparts, and depend explicitly on the cutoff scales. However, when the cutoff scale of the stochastic theory coincides with that of the deterministic theory, the equal-time correlators of the conserved densities become identical to those of the ``standard'' non-relativistic theory \cite{landau9} (replacing the rest mass density with the relativistic enthalpy density), and do not show any dependence on the overall cutoff parameter. This is due to a spontaneous cancellation mechanism that we do not fully understand yet, and which seems to be specific to equal-time correlators (it will surely break down in correlators computed at non-equal times). In an upcoming article, we will use the regularized information current to develop a full-fledged first-order theory for hydrodynamics fluctuations, and we will compute the two-point correlators at non-equal times. Finally, it would be interesting to investigate how the modified KMS symmetry discussed in \cite{Mullins:2023ott} affects the determination of the regularized information current investigated here. This would also be crucial in order to generalize the considerations made here to the case of nonlinear fluctuating systems.

\section*{Acknowledgements}

LG is partially supported by Vanderbilt's Seeding Success Grant. JN and NA are partially supported by the U.S. Department of Energy, Office of Science, Office for Nuclear Physics
under Award No.  DE-SC0023861. ES has received funding from the European Union’s Horizon Europe research and innovation program under the Marie Sk\l odowska-Curie grant agreement No. 101109747. LG would like to thank N. Abbasi for an instructive discussion.

\appendix

\section{Alternative expressions for the information current of electric conduction}\label{AAA}

Let us recall that $J^0=\varphi+\tau \partial_t \varphi$ and $J^j=-D\partial^j \varphi+\Sigma \mathcal{E}^j$. Then, our alternative information current and entropy production rate for an electrically conducting Ohmic medium can be expressed as follows
\begin{equation}
    \begin{split}
E^0={}& \dfrac{1}{2} \bigg[ (J^0)^2+\dfrac{\tau}{D} J^j J_j+ \dfrac{\Sigma}{D} \mathcal{E}^j \mathcal{E}_j + \dfrac{\Sigma}{D} (1+\tau \Sigma)\mathcal{B}^j\mathcal{B}_j \bigg] \, , \\
E^j={}& J^0 J^j + \dfrac{\Sigma}{D} \big[(\mathcal{E}+\tau J) \times \mathcal{B} \big]^j \, , \\
\sigma ={}&  \dfrac{1+\tau \Sigma}{D} J^j J_j \, . \\
    \end{split}
\end{equation}
In fact, using the above constitutive relations for $J^\mu$, the equation $\partial_\mu E^\mu+\sigma=0$ can be recast in the following form:
\begin{equation}
    J^0 \partial_\mu J^\mu +\dfrac{\Sigma}{D} (\mathcal{E}^j{+}\tau J^j)\big[ \partial_t \mathcal{E}_j -(\nabla {\times} \mathcal{B})_j +J_j \big] +\dfrac{\Sigma}{D} (1{+}\tau \Sigma)\mathcal{B}^j \big[\partial_t \mathcal{B}_j + (\nabla {\times} \mathcal{E})_j \big] =0 \, ,
\end{equation}
which is automatically respected along all solutions of the equations of motion \eqref{maxone}. With a bit of algebra, one can also derive the conditions for timelike future directedness, which (not surprisingly) coincide with the conditions for causality and stability of the equations of motion.

\section{Quick derivation of Debye screening}\label{debby}

If we place a unit point charge at rest in a conducting medium, and we assume that such charge interacts with the medium only through electromagnetic interactions, then the Maxwell equations \eqref{maxone} are unchanged, except for the first one, which becomes $\partial_j \mathcal{E}^j=\varphi+\tau \partial_t \varphi+\delta^3(\textbf{x}-\textbf{y})$, where $\textbf{y}$ is the location of the point charge. Then, taking the divergence of the last equation of \eqref{maxone}, we obtain the following equation of motion for $\varphi$:
\begin{equation}
    \tau \partial^2_t \varphi +(1+\tau \Sigma )\partial_t \varphi +\Sigma \varphi -D \partial_j \partial^j \varphi =-\Sigma \delta^3(\textbf{x}-\textbf{y}) \, .
\end{equation}
This dynamical equation admits a stationary solution of the form
\begin{equation}\label{fourfour}
    \varphi(\textbf{x}) = -\int \dfrac{d^3 k}{(2\pi)^3} \dfrac{\Sigma \, e^{i\textbf{k}\cdot (\textbf{x}-\textbf{y})}}{\Sigma + D k^j k_j} \, .
\end{equation}
Considering that, in the stationary limit, the total charge density is $J^0=\varphi +\delta^3(\textbf{x}-\textbf{y})$, we can evaluate the Fourier integral in equation \eqref{fourfour} analytically, and we finally obtain an expression for $J^0$:
\begin{equation}
    J^0(\textbf{x})=\delta^3(\textbf{x}-\textbf{y}) - \dfrac{\Sigma e^{-\sqrt{\frac{\Sigma}{D}}|\textbf{x}-\textbf{y}|}}{4\pi D |\textbf{x}-\textbf{y}|} \, ,
\end{equation}
which coincides with the right-hand side of equation \eqref{gumpfone}. This equation tells us that the point charge surrounds itself with a shell of opposite charge, which completely screens its electric field over the length scale $\sqrt{D/\Sigma}$ -- this is the physics of Debye screening in plasmas.

\section{Computation of the information current for chiral hydrodynamics}\label{twonicks}

Following conventional methods \cite{GavassinoGibbs2021,GavassinoGENERIC2022} for evaluating the information current, we take the chiral fluid to be in weak thermal contact with a heat bath $H$. Here ``weak contact'' means that the value of every extensive quantity $Q_\text{tot}$ of the fluid+bath system is equal to the value $Q$ in the fluid plus the value $Q_H$ in the heat bath. The interactions between the fluid and the bath conserve total energy-momentum $P^\nu + P^\nu_H$ and particle number $N + N_H$. The relevant thermodynamic potential is
\begin{equation} \label{chir-potential}
    \Phi = S + \alpha^\star N + \beta^\star_\nu P^\nu\, ,
\end{equation}
where $S$ is the fluid's entropy \cite{GavassinoTermometri2020}. The coefficients $\alpha^\star=-\frac{\partial S_H}{\partial N_H}$ and $\beta^\star_\nu = - \frac{\partial S_H}{\partial P^\nu_H}$ are assumed to be constant properties of the bath, i.e., the bath is thermodynamically large. As a consequence of the second law of thermodynamics and charge conservation for the fluid+bath system, $\Phi$ is a non-decreasing function of time that attains its maximum when the system reaches global thermodynamic equilibrium.

To identify this global equilibrium state, we consider a one-parameter family of fluid states $\{T(h), \mu(h), u^\mu(h)\}$ parameterized by $h$ that reduces to global equilibrium when $h=0$. Here $T$, $\mu$, and $u^\mu$ are the temperature, chemical potential, and velocity fields, respectively. The spacetime dependence of all fields is suppressed in this appendix. Adopting the notation $\dot f = df/dh$, the maximality of $\Phi$ in equilibrium implies that every observer sees $\dot \Phi(h=0)=0$ for any choice of one-parameter family.

We now proceed to evaluate $\dot\Phi$. On a three-dimensional spacelike hypersurface $\Sigma$ with future-directed area element $d\Sigma_\mu$, the fluid's entropy, energy-momentum and particle number are given by $S=\int d\Sigma_\mu S^\mu$, $P^\nu = \int d\Sigma_\mu T^{\mu\nu}$, and $N=\int d\Sigma_\mu J^\mu$, so we can write
\begin{equation}
	\Phi = \int d\Sigma_\mu \phi^\mu\, , \qquad\qquad \phi^\mu = S^\mu + \alpha^\star J^\mu + \beta^\star_\nu T^{\mu\nu}.
\end{equation}
Inserting $S^\mu$, $J^\mu$ and $T^{\mu\nu}$ from \eqref{CHIR} gives
\begin{align}
    \phi^\mu &= [s + \alpha^\star n + \beta^\star_\nu u^\nu (\varepsilon + P)] u^\mu + P \beta^{\star\mu} + (\xi_S + \alpha^\star \xi_J)\omega^\mu\, , \\
    \label{dotphi}
    \dot\phi^\mu &= [\dot s + \alpha^\star \dot n + \beta_\nu^\star u^\nu (\dot \varepsilon + \dot P)+ \beta_\nu^\star \dot u^\nu (\varepsilon + P)] u^\mu + [s + \alpha^\star n + \beta^\star_\nu u^\nu (\varepsilon + P)]\dot u^\mu + \dot P \beta^{\star\mu} \\
    &\quad  + (\xi_S + \alpha^\star \xi_J)\dot\omega^\mu + (\dot \xi_S + \alpha^\star \dot \xi_J)\omega^\mu\, . \nonumber
\end{align}
Note that $\alpha^\star$ and $\beta^\star_\nu$, being properties of the bath, do not depend on $h$. \emph{Non-rotating} equilibria, to which we specialize immediately\footnote{In fact, we have effectively specialized already by neglecting to include angular momentum as a conserved quantity in the thermodynamic potential \eqref{chir-potential}. Including it would not modify our conclusions.}, have $T$, $\mu$, and $u^\mu$ constant over all spacetime, so the last term in \eqref{dotphi} vanishes at $h=0$. Furthermore, the second-to-last term vanishes at $h=0$ when integrated over any spacelike hypersurface $\Sigma$. To see this, note that when $h=0$ we have
\begin{align}
	\int d\Sigma_\mu \dot\omega^\mu &= \frac{1}{2}\int d\Sigma_\mu \epsilon^{\mu\alpha\beta\gamma}u_\alpha \partial_\beta \dot u_\gamma\, .
\end{align}
This evidently vanishes on hypersurfaces $\Sigma^{(0)}$ that satisfy $d\Sigma_\mu \propto u_\mu$. For any other spacelike hypersurface $\Sigma$, note that the union $\Sigma + \Sigma^{(0)}$ forms the boundary of some four-dimensional hypervolume $\Omega$ with infinite spatial extent. Then
\begin{align} \label{gauss}
    \begin{split}
        \int_{\Sigma} d\Sigma_\mu \left( \frac{1}{2} \epsilon^{\mu\alpha\beta\gamma} u_\alpha \partial_\beta \dot u_\gamma \right) &= \int_{\Sigma + \Sigma^{(0)}} d\Sigma_\mu \partial_\beta \left( \frac{1}{2} \epsilon^{\mu\alpha\beta\gamma} u_\alpha \dot u_\gamma \right) \\
	&= \int d\Omega\ \partial_\mu \partial_\beta \left( \frac{1}{2}\epsilon^{\mu\alpha\beta\gamma}u_\alpha \dot u_\gamma \right) \\
	&= 0\, ,
    \end{split}
\end{align}
as claimed. To obtain the second equality, we used Gauss' law under the assumption that $\dot u^\mu$ vanishes sufficiently fast at spatial infinity. It is easily verified that the remaining terms in $\dot\phi^\mu(h=0)$ vanish if and only if the equilibrium state satisfies
\begin{align} \label{chiral-eqm}
	\mu/T = \alpha^\star \qquad \qquad u_\nu/T = \beta_\nu^\star\, .
\end{align}

The (conventional, unregulated) information current is defined by
\begin{align}
    E^\mu = \phi^\mu(0) - \phi^\mu(h).
\end{align}
To obtain its quadratic approximation, we expand $\phi^\mu(h) = \phi^\mu(0) + \frac{1}{2}h^2\ddot \phi^\mu(0) + \mathcal O(h^3)$. By differentiating \eqref{dotphi} once more and then inserting \eqref{chiral-eqm}, we find
\begin{align} \label{E-chir}
    T\ddot \phi^\mu(0) &= -[\dot T \dot s + \dot \mu \dot n + \dot u_\nu \dot u^\nu (\varepsilon + P)]u^\mu - 2\dot P\dot u^\mu + (T\xi_S + \mu\xi_J) \ddot \omega^\mu + 2 (T\dot \xi_S + \mu \dot\xi_J) \dot\omega^\mu\, .
\end{align}
Here we have used $u_\mu \dot u^\mu = \dot u_\mu \dot u^\mu + u_\mu \ddot u^\mu = 0$, which follows from differentiating the normalization $u_\mu u^\mu$, as well as the first law of thermodynamics $\dot \varepsilon = T \dot s + \mu \dot n$ and the Euler relation $Ts+\mu n = \varepsilon + P$. Equation \eqref{E-chir} can be simplified further by inserting the definition of $\bar\xi$ in \eqref{xibar} and applying the constraint $T \dot \xi_S + \mu \dot \xi_J = \frac{4\bar\xi}{\varepsilon + P}\dot P$, which follows from a first-order entropy-current analysis \cite{Son:2009tf, Stephanov:2015roa}. We also note that the term $\bar\xi\epsilon^{\mu\alpha\beta\gamma}u_\alpha \partial_\beta \ddot u_\gamma$ contained within the second-to-last term of \eqref{E-chir} vanishes when integrated over any spacelike hypersurface for the same reason as in \eqref{gauss}. Finally, adopting the conventional notation $\delta f = f(h) - f(0)$, we have
\begin{align}
    \begin{split}
        TE^\mu &= - \frac{1}{2}h^2 T\ddot \phi^\mu(0) \\
        &= \frac{1}{2}[\delta T \delta s + \delta \mu \delta n + \delta u^\nu \delta u_\nu(\varepsilon+ P)] u^\mu + \delta P \delta u^\mu - \frac{2\bar\xi}{\varepsilon + P}\delta P \epsilon^{\mu\alpha\beta\gamma} u_\alpha \partial_\beta \delta u_\gamma - \bar\xi \epsilon^{\mu\alpha\beta\gamma}\delta u_\alpha \partial_\beta \delta u_\gamma\, .
    \end{split}
\end{align}
When evaluated in the local rest frame, where $u^\mu = (1, \vec 0)$ and $\delta u^\mu = (0, \vec{\delta u})$, this reduces to \eqref{grzi}.

\bibliography{Biblio}

\label{lastpage}

\end{document}